%% file: paper.tex
\documentclass[fleqn,usenatbib]{mnras}

\usepackage[T1]{fontenc}
\usepackage[utf8]{inputenc}
\usepackage{microtype}
\usepackage{amsmath}

\usepackage{txfonts}

\usepackage{mathtools}	% Advanced maths commands
\usepackage{bm}
\usepackage{pgf}
\usepackage[draft,footnote,nomargin]{fixme}
\usepackage{cleveref}

\fxsetup{theme=color,mode=multiuser}
\FXRegisterAuthor{ch}{ach}{\color{fxnote}CH}
\FXRegisterAuthor{th}{ath}{\color{fxerror}TH}

\newcommand{\md}{\mathrm{d}}
\newcommand{\diff}{\mathrm{d}}
\newcommand{\vdiff}[1]{\diff\bm{#1}}

\newcommand{\fraction}[2]{\dfrac{\displaystyle#1}{\displaystyle#2}}

\newcommand{\pder}[2]{\fraction{\partial#1}{\partial#2}}

\DeclareFontFamily{U}{mathb}{\hyphenchar\font45}
\DeclareFontShape{U}{mathb}{m}{n}{
      <5> <6> <7> <8> <9> <10> gen * mathb
      <10.95> mathb10 <12> <14.4> <17.28> <20.74> <24.88> mathb12
      }{}
\DeclareSymbolFont{mathb}{U}{mathb}{m}{n}
\DeclareFontSubstitution{U}{mathb}{m}{n}
\DeclareMathSymbol{\curvearrowbotright}{3}{mathb}{"F4}

\newcommand{\gobelow}{\curvearrowbotright}

\newcommand{\ext}{\mathrm{ext}}
\newcommand{\tot}{\mathrm{tot}}
\newcommand{\prtcl}{\mathrm{p}}
\newcommand{\kJ}{k_{\mathrm{J}}}

\renewcommand{\Im}{\operatorname{Im}}

 \title[Linear response of stellar systems]{The linear response of stellar systems does not diverge at marginal stability}

\author[C. Hamilton \& T. Heinemann]{
  Chris Hamilton$^{1}$\thanks{E-mail: chamilton@ias.edu} and Tobias Heinemann$^{2}$\\
$^1$Institute for Advanced Study, Einstein Drive, Princeton, NJ 08540, USA\\
$^2$Niels Bohr International Academy, Niels Bohr Institute, Blegdamsvej 17, 2100 Copenhagen, Denmark}

\pubyear{2022}

\begin{document}
\label{firstpage}
\pagerange{\pageref{firstpage}--\pageref{lastpage}}
\maketitle

%%%%%%%%%%%%%%
%%%%%%%%%%%%%%
%%%%%%%%%%%%%%
%%%%%%%%%%%%%%
%%%%%%%%%%%%%%%hhgiie  spol 

%%%%%%%%%%%%%%
%%%%%%%%%%%%%%
%%%%%%%%%%%%%%

\begin{abstract}
The linear response of a stellar system's gravitational potential to a perturbing mass
comprises
two distinct contributions.  
Most famously, the system will respond by forming a polarization `wake' around the perturber.
 At the same time, the perturber may also excite one or more
`Landau modes', i.e. coherent oscillations of the entire stellar system which are either stable or unstable 
depending on the system parameters.
The amplitude of the first (wake) contribution
is known to diverge as a system approaches marginal stability.
In this paper we consider the linear response of a homogeneous stellar system
to a point mass moving on a straight line orbit.
We prove analytically that the divergence of the wake response is in fact cancelled by a corresponding divergence in the Landau mode response,
rendering the total response finite.
We demonstrate this cancellation explicitly for a box of stars with Maxwellian velocity distribution.
Our results imply that polarization wakes may be much less efficient drivers of secular evolution than previously thought.
More generally,
any prior calculation that accounted for wakes but ignored modes {--- such as those based on the Balescu-Lenard equation ---} may need to be revised.
\end{abstract}


%%%%%%%%%%%%%%%%%%%%%%%%%%%%%%%%%%%%%%%%%%%%%%%%%%

%%%%%%%%%%%%%%%%% BODY OF PAPER %%%%%%%%%%%%%%%%%%

%%%%%%%%%%%%%%
%%%%%%%%%%%%%%
%%%%%%%%%%%%%%
%%%%%%%%%%%%%%
%%%%%%%%%%%%%%
%%%%%%%%%%%%%%
%%%%%%%%%%%%%%
%%%%%%%%%%%%%%

%%%%%%%%%%%%%%%%%%%%%%%%%%%%%%%%%%%%%%%%%%%%%%%%%%%%%%%%%%%%
\section{Introduction}
\label{sec:Introduction}
%%%%%%%%%%%%%%%%%%%%%%%%%%%%%%%%%%%%%%%%%%%%%%%%%%%%%%%%%%%%

Many of the crowning achievements of galactic dynamics --- 
such as our understanding of spiral structure \citep{lin1964spiral,julian1966non,Lynden-Bell1972-ve,Sellwood2014-xo}, of dynamical friction \citep{chandrasekhar1943dynamical}, of galaxy stability \citep{palmer1994stability}, and of bar-halo interactions \citep{Tremaine1984-wt, Weinberg1985-en} --- are underpinned, explicitly or otherwise, by the theory of linear response.
At the most basic level, linear response theory answers the question:
how does a stellar system respond to a weak gravitational perturbation?
This perturbation can take various forms; depending on the context, bars, spiral modes, molecular clouds, infalling satellites, dark matter substructure, and internal Poisson noise all constitute legitimate perturbers of a galaxy \citep{Weinberg2001-ok,Pichon2006-ak}.
Though most of the conclusions we draw in this paper will be valid for systems undergoing arbitrary perturbations,
for concreteness and concision we will focus here on systems perturbed by a single point mass.

The foundational study of a stellar system perturbed by a heavy point mass is \cite{chandrasekhar1943dynamical}'s paper on dynamical friction.
Chandrasekhar showed that, to linear order in the perturber mass, the stellar system responds by forming a `wake' behind the perturber.
This wake is said to `polarize' or `dress' the gravitational potential of the perturber, typically increasing its effective mass, in a manner analogous to Debye shielding in electrostatic plasma \citep{Ichimaru1973}.
The second-order effect is that the gravitational field of the wake backreacts on the perturber, producing an effective frictional drag.
Chandrasekhar's picture of wake-drive friction now underlies huge swathes of astrophysical theory concerning galaxy mergers, supermassive black hole coalescence, the nature of dark matter, and so on (e.g. \citealt{begelman1980massive,boylan2008dynamical,Lancaster2020-vt}).

Moreover, several authors have recognized that the linear, wake-like response 
described above is particularly prominent if the stellar system in question is only \textit{weakly stable}. 
By `stable' here we mean that the
Landau modes --- i.e. the set of self-consistent, coherent oscillations that the system can support in the absence of a perturber --- are all exponentially damped; the word `weakly' means that the associated damping timescale is long (in a sense we define more precisely below).
The key point is that in weakly stable systems, 
the wake response can be strongly amplified by its own self-gravity.

This amplification is important for the dynamical friction experienced by a sinking satellite \citep{Weinberg1989-cj}.  
It also has a profound effect upon 
the response of galactic disks to co-orbiting perturbers like molecular clouds \citep{d2013self,Sellwood2021-rt}.
For instance, in a classic analysis, \cite{julian1966non}
inserted a point mass into a local model of a stable, shearing stellar disk.
Working in the time-asymptotic limit, they found that the
wake induced by the point mass 
was much more massive in disks with smaller Toomre $Q$ (see their Table 1; also \citealt{binney2020shearing}). In other words, the less stable they made the disk, the more strongly amplified was the response. 
These self-amplified shearing wakes are still one of the primary explanations offered for the existence of 
spiral structure in galaxies.
Naively, however, one might expect that as the system is driven towards marginal stability ($Q\to 1$) the linear 
wake response will diverge.

Similarly, and importantly for this paper, 
\cite{weinberg1993nonlocal} {and \cite{magorrian2021stellar} have}
studied the response of a \textit{homogeneous, periodic}
stellar system to a point mass perturber moving on a straight-line orbit.\footnote{{Their approaches to the linear response calculation differed. Weinberg used Laplace transforms in time, like we do here.
Magorrian opted to work in the time domain but performed a Fourier-transform in velocity space; he also allowed his perturber to grow adiabatically in mass between $t=-\infty$ to $t=0$.}} 
{Both authors} derived an expression for the linear wake-like
perturbation to the system's 
distribution function in the time-asymptotic limit (i.e. after waiting a time long compared to the damping time of all Landau modes)
---
see eq. (22) of \citet{weinberg1993nonlocal}, and eq. (40) of \cite{magorrian2021stellar}.
However, {these expressions} manifestly diverge if 
one of the system's Landau mode frequencies tends towards the
(real) frequency of the perturbation, 
which is 
possible as the system approaches marginal stability.
%Thus, one could imagine placing a perturber into a hot, stable stellar system, and gradually cooling the system down until it was on the verge of instability. 
%Based on the above considerations, one would expect the perturber to induce a divergent wake-like response in this case.
%\cite{weinberg1993nonlocal} also used his linear expression to derive a kinetic equation for the slow evolution of the stellar system's spatially-averaged distribution function. This again diverged as the system approached marginal stability.
More broadly, {many `quasilinear'} kinetic theories that purport to describe the 
secular evolution of stellar systems predict a rate of evolution that is proportional to the product of two linear wake-like terms (e.g. \citealt{Binney1988-zy,Fouvry2018-gi}, {and eq. (79) of \citealt{magorrian2021stellar})}.
It has been claimed that this rate will diverge as the system approaches marginal stability, and
this behavior has been likened to the phenomenon of critical opalescence
(e.g. \citealt{chavanis2023secular}).

These divergent wake results are worrying,
since they predict that near marginal stability, 
the linear response of stellar systems can become arbitrarily (and hence nonlinearly) large.
Taken at face value, this suggests that linear and quasilinear theories are
incapable of describing accurately the dynamics of weakly stable stellar systems.
%
%like the Galactic disk \citep{rafikov2001local}.
%
%and found that the amplitude of the wake-like response diverges for wavenumbers $k$ approaching the Jeans wavenumber $k_\mathrm{J}$.
%Said differently, these authors were all considering
%stable stellar systems which harbored a Landau mode with some growth rate $\gamma < 0$ \citep{Binney2008-ou}; in each case, they found that the wake-like response to the perturber was amplified indefinitely as they changed the parameters of their system such that $\gamma \to 0^-$. 
%
{However, the divergences can all be traced back to an erroneous time-asymptotic 
assumption, in a regime where the required timescale separation does not actually exist.
Precisely, the above authors were (explicitly or otherwise)
assuming that all the Landau modes of the stellar system had decayed.
But as a system approaches marginal stability, the timescale for this decay 
becomes infinitely long.}
Said differently, {near marginal stability,}
a wake calculation on its own is an incomplete description of the finite-time linear response:
one should also account directly for the contribution of the system's Landau modes. 
As is well-known in plasma theory, this Landau mode contribution \textit{also} diverges as a system is driven towards marginal stability
(e.g. \citealt{Rogister1968-tb, hatori1969nonstationary,oberman1970advanced}). The key point is that \textit{the divergence in the Landau mode contribution cancels the divergence in the wake contribution}, such that the total linear response --- the only 
physically relevant quantity --- is rendered finite.
Not only is the divergence cured, but the amplitude of the total linear response is much smaller than one would naively predict based on the wake calculation alone, meaning linear theory need not necessarily be abandoned.

The purpose of this paper is to demonstrate this basic fact in a simple stellar-dynamical context.
To this end, we consider an initially unperturbed, homogeneous box of stars, and calculate the linear response of its gravitational potential to an externally-imposed point-mass perturber moving on a straight-line orbit.
We include the Landau mode contribution to the response, and thereby prove that the total response is always finite regardless of the system's stability properties.
We also demonstrate the cancellation of the wake and mode divergences
explicitly for a box of stars with initially Maxwellian velocity distribution.

We present our calculations in \S\ref{sec:Linear_Theory}, and summarise in \S\ref{sec:Summary}.

%%%%%%%%%%%%%%%%%%%%%%%%%%%%%%%%%%%%%%%%%%%%%%%%%%%%%%%%%%%%
\section{Linear response theory}
\label{sec:Linear_Theory}
%%%%%%%%%%%%%%%%%%%%%%%%%%%%%%%%%%%%%%%%%%%%%%%%%%%%%%%%%%%%

%In this section we will solve the lineariised Vlasov-Poisson system
%as an initial value problem, accounting explicitly for the potential 
%fluctuations due to (i) wakes and (ii) Landau modes.
%For simplicity, we will start with the case where there is a perturbation in the initial distribution function and there are no external forces (\S\S\ref{sec:Setup}-\ref{sec:total_potential_fluctuation}).
%We then briefly mention the minor modifications needed to generalise the theory 
%to include external perturbations (\S\ref{sec:External}).

%\subsection{Lineariised Vlasov-Poisson system}
%\label{sec:Setup}

We consider a system of equal-mass stars interacting via Newtonian gravity and subject to an external potential force. Let $f(\bm{r},\bm{v},t)$ be the distribution function (DF), which we normalize in such a way that the average mass contained inside the infinitesimal phase-space
volume element $\vdiff{r}\vdiff{v}$ around the point $(\bm{r},\bm{v})$ at time $t$ is equal to
$f(\bm{r},\bm{v},t)\vdiff{r}\vdiff{v}$. We will assume that the system is collisionless, in which case the DF obeys the Vlasov equation
\begin{equation}
  \label{eq:vlasov}
  \pder{f}{t} + \bm{v}\cdot\pder{f}{\bm{r}}
  - \bm{\nabla}\Phi^{\tot}\cdot\pder{f}{\bm{v}} = 0.
\end{equation}
The total potential $\Phi^{\tot}$ consists of two parts,
\begin{equation}
  \label{eq:total-potential}
  \Phi^{\tot}(\bm{r},t) = \Phi(\bm{r},t) + \Phi^{\ext}(\bm{r},t).
\end{equation}
The internal potential $\Phi$ satisfies the Poisson equation
\begin{equation}
  \label{eq:poisson}
  \nabla^2\Phi = 4\pi G\!\int\!\vdiff{v}\,f.
\end{equation}
The external potential $\Phi^{\ext}$ will be left unspecified for now.

Let us assume further that our system is confined to a periodic box {of} volume $V$. Then we can
develop any function on phase space as a Fourier series: for instance, 
\begin{equation}
  \label{eq:fourier-series}
  f (\bm{r},\bm{v},t)
  = \sum_{\bm{k}}\exp(i\bm{k}\cdot\bm{r})
  f_{\bm{k}}(\bm{v},t),
\end{equation}
where {the wavenumbers $\bm{k}$ have components $k_i = 2\pi n_i/L_i$ where $n_i$ is an integer and $L_i$ is the length of the $i$th side of the box}\footnote{{Our results also hold in the special case of infinite homogeneous systems $V\to\infty$, by replacing $V^{-1}\sum_{\bm{k}}\to (2\pi)^{-3}\int \md \bm{k}$.}}, and 
\begin{equation}
  \label{eq:fourier-coefficient}
  f_{\bm{k}}(\bm{v},t)
  = \frac{1}{V}\int\md \bm{r}\exp(-i\bm{k}\cdot\bm{r})
  f (\bm{r},\bm{v},t),
\end{equation}
are the spatial Fourier coefficients. 
We may also define the Laplace transform of these coefficients as
\begin{equation}
  \tilde{f}_{\bm{k}}(\bm{v},z)
  = \int_0^\infty\diff t\exp(izt)
  f_{\bm{k}}(\bm{v},t).
\end{equation}
This is defined for all complex frequencies $z$ with $\Im z\ge\eta$, where $\eta$
is larger than the growth rate of the most unstable Landau mode of the system (see below) or, if the
system is stable, larger than zero.

{We assume a spatially uniform background.  Further, we assume}
that the fluctuations in the DF are small compared to, 
and {evolve} much more rapidly than,
the spatial average $f_0$, i.e.\ $f_{\bm{k}}\ll f_0$ for $\bm{k}\ne \bm{0}$.
Then we may linearise the Vlasov equation \labelcref{eq:vlasov}
and Fourier-Laplace transform the result to arrive at 
\begin{equation}
  \label{eq:vlasov-laplace}
  i(\bm{k}\cdot\bm{v} - z)
  \tilde{f}_{\bm{k}}(\bm{v},z)
  - i\bm{k}\cdot\pder{f_0 }{\bm{v}}\tilde{\Phi}^\mathrm{tot}_{\bm{k}}(z)
  = f_{\bm{k}}(\bm{v},0).
\end{equation}
To simplify matters, we will assume that at $t=0$, there are no fluctuations in the DF,
i.e.\ $f_{\bm{k}}(\bm{v},0) = 0$.
The initial internal potential fluctuation is then just $\Phi_{\bm{k}}(0) = 0$, so that
$\Phi_{\bm{k}}^\mathrm{tot}(0) = \Phi^\mathrm{ext}_{\bm{k}}(0)$.
Solving \cref{eq:vlasov-laplace} for $\tilde{f}_{\bm{k}}(\bm{v},z)$ and inserting
the result into the linearized Poisson equation
\begin{equation}
    \label{eq:poisson-laplace}
    -k^2\tilde{\Phi}_{\bm{k}}(z) = 4\pi G\!\int\!\vdiff{v}\,\tilde{f}_{\bm{k}}(\bm{v},z)
\end{equation}
then yields after a little bit of algebra the equation
\begin{equation}
    \epsilon_{\bm{k}}(z)\tilde{\Phi}_{\bm{k}}^{\tot}(z)
    = \tilde{\Phi}_{\bm{k}}^{\ext}(z),
    \label{eqn:vlasov-poisson}
\end{equation}
where
\begin{equation}
  \label{eq:permittivity}
  \epsilon_{\bm{k}}(z)
  = 1 + \frac{4\pi G}{k^2}\!\int\!\vdiff{v}\,
  \frac{1}{\bm{k}\cdot\bm{v} - z}\,
  \bm{k}\cdot\pder{f_0 }{\bm{v}}
  \qquad(\Im z > 0)
\end{equation}
is the \emph{dispersion function}, which for $\Im z\le 0$ is defined through analytic continuation (see below).
The zeroes of $\epsilon_{\bm{k}}(z)$ correspond to the frequencies of the Landau modes that the system supports.

The total gravitational potential {in the time domain} is given by {the inverse Laplace transform:}
\begin{equation}
    \label{eq:inverse-laplace}
    \Phi_{\bm{k}}^{\tot}(t)
    = \int_{i\eta-\infty}^{i\eta+\infty}\frac{\diff z}{2\pi}\exp(-izt)
    \tilde{\Phi}_{\bm{k}}^{\tot}(z).
\end{equation}
As we remarked above, $\eta$ is larger than the growth rate of the most unstable Landau mode or,
if the system is stable, larger than zero. This means that by construction,
$\epsilon_{\bm{k}}(z)\ne0$ along the integration contour in \cref{eq:inverse-laplace}. From \cref{eqn:vlasov-poisson} it thus follows that
\begin{equation}
    \label{eq:total-potential-IL}
    \Phi_{\bm{k}}^{\tot}(t)
    = \int_{i\eta-\infty}^{i\eta+\infty}\frac{\diff z}{2\pi}\exp(-izt)
    \frac{\tilde{\Phi}_{\bm{k}}^{\ext}(z)}{\epsilon_{\bm{k}}(z)}.
\end{equation}

\subsection{Response of an unperturbed system to a point mass moving at constant velocity}

Suppose $ \Phi^{\ext}$ is generated by an externally-imposed point mass $m$
moving on a straight-line trajectory $\bm{r}=\bm{r}_{\prtcl} + \bm{v}_{\prtcl}t$ for constant
$\bm{r}_{\prtcl}$ and $\bm{v}_{\prtcl}$.
Then the external potential $\Phi^\mathrm{ext}$ satisfies
\begin{align}
  \label{eq:poisson-ext}
  \nabla^2 \Phi^{\ext} = 4\pi Gm\,\delta(\bm{r} - \bm{r}_{\prtcl} - \bm{v}_{\prtcl} t).
\end{align}
From this it follows that the Fourier coefficients of the external potential
are
\begin{equation}
  \label{eq:point-mass-potential-IL}
  \Phi_{\bm{k}}^{\ext}(t)
  = \exp(-i\bm{k}\cdot\bm{v}_{\prtcl}t)\Phi_{\bm{k}}^{\ext}(0),
\end{equation}
where
\begin{equation}
  \Phi_{\bm{k}}^{\ext}(0)
  = -\frac{4\pi G}{k^2}\frac{m}{V}\exp(-i\bm{k}\cdot\bm{r}_{\prtcl}).
\end{equation}
The Laplace transform of eq. \eqref{eq:point-mass-potential-IL} is
\begin{equation}
  \label{eq:point-mass-potential}
  \tilde{\Phi}_{\bm{k}}^{\ext}(z)
  = \frac{\Phi_{\bm{k}}^{\ext}(0)}{i(\bm{k}\cdot\bm{v}_{\prtcl} - z)}.
\end{equation}
Inserting this into \cref{eq:total-potential-IL} yields the total gravitational potential in the form of
\begin{equation}
    \label{eq:total-potential-IL-point-mass}
    \Phi_{\bm{k}}^{\tot}(t)
    = \int_{i\eta-\infty}^{i\eta+\infty}\frac{\diff z}{2\pi i}
    \frac{\exp(-izt)}{(\bm{k}\cdot\bm{v}_{\prtcl} - z)\epsilon_{\bm{k}}(z)}
    \Phi_{\bm{k}}^{\ext}(0).
\end{equation}

\subsection{The response at late times}

The integral in \cref{eq:total-potential-IL-point-mass} can be carried out by 
closing the integration contour with a very large semicircle in the lower-half plane on which the integrand vanishes
\citep{thorne2017modern}.
Provided that the integrand's only singularities are poles, the inverse Laplace transform is then equal to $2\pi i$ times the sum of residues of the integrand at those poles.%\thnote{I think it would really help if we added a small figure that illustrates this.}

In order to follow this procedure, we must analytically continue the dispersion function $\epsilon_{\bm{k}}(z)$, defined in \cref{eq:permittivity} for $\Im z>0$, into the lower half of the complex plane.
This is achieved by taking the integral over the parallel component of the stellar velocity $\bm{k}\cdot\bm{v}/k$ along the so-called Landau contour, which is deformed in such a way that it always passes below the point $\bm{k}\cdot{\bm{v}}=z$ (e.g. \citealt{Binney2008-ou}).
We denote this by
\begin{equation}
  \label{eq:permittivity-plus}
  \epsilon_{\bm{k}}(z)
  = 1 + \frac{4\pi G}{k^2}\!\int\!\vdiff{v}\,
  \frac{1}{(\bm{k}\cdot\bm{v} - z)^\gobelow}\,
  \bm{k}\cdot\pder{f_0 }{\bm{v}},
\end{equation}
which, unlike \cref{eq:permittivity} is defined for all $z$.

The dispersion function as given in \cref{eq:permittivity-plus} is analytic in the whole complex plane,
so its zeroes (i.e.\ the Landau mode frequencies) are all isolated\textbf{\footnote{{Technically, $\epsilon_{\bm{k}}(z)$ 
can also exhibit other unusual features, e.g. branch cuts 
associated with singularities of $f_0(\bm{v})$ for complex $\bm{v}$.
These give rise to behavior different from either the wakes or Landau modes considered here 
(e.g. \citealt{lee2023cauchy}).  We ignore these complications.
}}}.
Let us assume first that the system is stable, which means that all Landau mode frequencies are located in the lower half of the complex plane.
The exponential factor in \cref{eq:total-potential-IL-point-mass} then ensures that the contributions from the residue at each Landau mode frequency eventually decays. Asymptotically, the only non-zero contribution comes from the residue at $z=\bm{k}\cdot\bm{v}_{\prtcl}$, which is given by
\begin{equation}
    \label{eq:response-no-eigenmodes}
    \Phi_{\bm{k}}^{\tot}(t)
    = \frac{1}{\epsilon_{\bm{k}}(\bm{k}\cdot\bm{v}_{\prtcl})}\Phi_{\bm{k}}^{\ext}(t)
    \quad\text{for}\quad t\gg-1/\gamma_{\bm{k}},
\end{equation}
where $\gamma_{\bm{k}}<0$ is the growth rate of the most weakly damped mode (i.e. all other modes have imaginary parts which are more negative than $\gamma_{\bm{k}}$).
{Typically one interprets \eqref{eq:response-no-eigenmodes}} by saying that the external potential $\Phi^\mathrm{ext}_{\bm{k}}(t)$ has been `dressed' by collective effects, which are encapsulated in 
the factor $1/\epsilon_{\bm{k}}(\bm{k}\cdot\bm{v}_\mathrm{p})$.
The part of \eqref{eq:response-no-eigenmodes} which comes from the perturbed distribution of stars --- i.e. the right hand side of \eqref{eq:response-no-eigenmodes} minus $\Phi^\mathrm{ext}_{\bm{k}}(t)$ --- is typically referred to as the `wake'.
As a shorthand,
in the following subsection we will refer to the entire right hand side of \eqref{eq:response-no-eigenmodes} as the `wake' contribution to the potential.

The dressed potential as given in \cref{eq:response-no-eigenmodes} clearly
diverges if $\epsilon_{\bm{k}}(z)$ has a zero at the real frequency $z=\bm{k}\cdot\bm{v}_\mathrm{p}$.
A necessary condition for this is that the system is marginally stable, 
i.e. $\gamma_{\bm{k}}=0$.
Put differently, if one considers a marginally stable system, one can always choose a perturber velocity 
$\bm{v}_\mathrm{p}$ such that 
the right hand side of \eqref{eq:response-no-eigenmodes} is infinite. 
%Of course, the condition $t\gg-1/\gamma_{\bm{k}}$ is in conflict with the condition of marginal stability. 
%This is most easily seen by supposing that a) the ``external star''\thnote{Need a better name for this.} is either at rest or that the wave number $\bm{k}$ is perpendicular to its velocity $\bm{v}_{\prtcl}$, and b) the most weakly damped eigenmode has a purely imaginary frequency $z_{\bm{k}}=i\gamma_{\bm{k}}$ (as is the case for a Maxwellian stellar system).\thnote{More generally, this is true for the kappa distribution
%\begin{equation}
%    f(\bm{v})
%    = \frac{\rho}{(2\pi)^{3/2}\sigma^3}
%    \frac{\Gamma(\kappa + 1)}{\kappa^{3/2}\Gamma(\kappa - 1/2)}
%    \frac{1}{[1 + v^2/(2\kappa\sigma^2)]^{1+\kappa}},
%\end{equation}
%which includes both the Maxwellian (in the limit $\kappa\to\infty$) and the Lorentzian ($\kappa=1$).}
%Under these conditions, \cref{eq:response-no-eigenmodes} evidently diverges as the system approaches marginal stability, i.e.\ as $\gamma_{\bm{k}}\to0$.
The key point of our paper is that the divergence of \eqref{eq:response-no-eigenmodes} as the system approaches 
marginal stability is spurious, and 
stems from the neglect of the Landau mode potential itself --- or, equivalently, from the erroneous assumption that one can always wait a time $t$ long compared to $-1/\gamma_{\bm{k}}$ such that \eqref{eq:response-no-eigenmodes} becomes valid.  No such $t$ exists for marginally stable systems, and even for very weakly stable systems, $-1/\gamma_{\bm{k}}$ may be comparable to other timescales of interest such as the relaxation time, rendering \eqref{eq:response-no-eigenmodes} invalid for practical purposes.

To cure the divergence, let us assume that the frequency of the system's most weakly damped mode is a single, simple zero of $\epsilon_{\bm{k}}(z)$.\footnote{The symmetry
$\epsilon_{\bm{k}}(z)^\ast=\epsilon_{\bm{k}}(-z^\ast)$
implies that if $\omega_{\bm{k}}+i\gamma_{\bm{k}}$ is a Landau mode frequency, then so is
$-\omega_{\bm{k}}+i\gamma_{\bm{k}}$. Thus, in general, there is not a single most weakly damped mode at each $\bm{k}$, but rather
a pair of most weakly damped modes. 
However, an exception to this occurs if $\omega_{\bm{k}}=0$, in which case there \textit{is}
only a single most weakly damped mode.
%In a Maxwellian stellar system, the most weakly damped mode has precisely this property.  
This is true for the kappa distribution
\begin{equation}
    f(\bm{v})
    = \frac{\rho}{(2\pi)^{3/2}\sigma^3}
    \frac{\Gamma(\kappa + 1)}{\kappa^{3/2}\Gamma(\kappa - 1/2)}
    \frac{1}{[1 + v^2/(2\kappa\sigma^2)]^{1+\kappa}},
    \label{eqn:kappa}
\end{equation}
which includes both the Maxwellian ($\kappa\to\infty$) and the Lorentzian ($\kappa=1$) as limiting cases.
The zeroes of $\epsilon_{\bm{k}}(z)$ for these DFs are also all simple.
Since these are the cases of most interest to us, we will assume a single, simple mode hereafter; 
one can straightforwardly generalize our treatment further if desirable \citep{oberman1970advanced}.}
%Under these conditions, \cref{eq:response-no-eigenmodes} evidently diverges as the system approaches marginal stability, i.e.\ as $\gamma_{\bm{k}}\to0$.
This means that, without loss of generality, we can write
\begin{equation}
    \label{eq:simple-zero}
    \epsilon_{\bm{k}}(z) = (z - z_{\bm{k}})\alpha_{\bm{k}}(z)
    \quad\text{with}\quad
    \alpha_{\bm{k}}(z_{\bm{k}})\ne 0,
\end{equation}
cf.\ \cite{Balescu1963-ye}. Note that $\epsilon_{\bm{k}}'(z_{\bm{k}})=\alpha_{\bm{k}}(z_{\bm{k}})$,
where $\epsilon_{\bm{k}}'(z)$ denotes the derivative of $\epsilon_{\bm{k}}(z)$ with respect to $z$.
We insert \cref{eq:simple-zero} into \cref{eq:total-potential-IL-point-mass}, and again close the contour with a large semicircle in the lower half plane, but this time we include the contribution both from the residue at $z=\bm{k}\cdot\bm{v}_{\prtcl}$ \textit{and} from the residue at $z=z_{\bm{k}}$.  The result is
\begin{equation}
    \label{eq:non-divergent-response}
    \Phi_{\bm{k}}^{\tot}(t)
    = \left[
        \frac{1}{\epsilon_{\bm{k}}(\bm{k}\cdot\bm{v}_{\prtcl})}
        - \frac{\exp\bigl(i(\bm{k}\cdot\bm{v}_{\prtcl} - z_{\bm{k}})t\bigr)}
        {(\bm{k}\cdot\bm{v}_{\prtcl} - z_{\bm{k}})\epsilon_{\bm{k}}'(z_{\bm{k}})}
    \right]
    \Phi_{\bm{k}}^{\ext}(t).
\end{equation}
This equation is valid for times $t\gg-1/\tilde{\gamma}_{\bm{k}}$, where $\tilde{\gamma}_{\bm{k}}$ is the growth rate of the \emph{second} most weakly damped mode.
The first term on the right hand side is the same as in \eqref{eq:response-no-eigenmodes}.  It represents the dressed potential of the perturber and, as we know, it diverges as $z_{\bm{k}}\to\bm{k}\cdot\bm{v}_{\prtcl}$. The second term is the potential of the Landau mode that the perturber has excited. It 
also diverges as $z_{\bm{k}}\to\bm{k}\cdot\bm{v}_{\prtcl}$, but it does so in a way that precisely cancels the divergence of the first term.
Indeed, letting $z_{\bm{k}}\to\bm{k}\cdot\bm{v}_{\prtcl}$ yields\footnote{Note that the time dependence of the second term on the right hand side of \eqref{eqn:total-potential-marginal-stability} is $t\exp(-iz_{\bm{k}}t)$. When carrying out the inverse Laplace transform using the calculus of residues, such a time-dependence is the signature of a second-order pole (see Appendix~A in \citet{Balescu1963}). This is sensible because the two simple poles at $z=\bm{k}\cdot\bm{v}_{\prtcl}$ and $z=z_{\bm{k}}$ merge to become a second-order pole as $z_{\bm{k}}\to\bm{k}\cdot\bm{v}_{\prtcl}$.}
%%%%%%%%%%%%%%%%%
\begin{equation}
    \lim_{z_{\bm{k}}\to\bm{k}\cdot\bm{v}_{\prtcl}}\Phi_{\bm{k}}^{\tot}(t)
    = -\left[
    \frac{\alpha_{\bm{k}}'(\bm{k}\cdot\bm{v}_\mathrm{p})}{\alpha_{\bm{k}}(\bm{k}\cdot\bm{v}_\mathrm{p})^2}
    + \frac{it}{\alpha_{\bm{k}}(\bm{k}\cdot\bm{v}_\mathrm{p})}
    \right]
    \Phi_{\bm{k}}^{\ext}(t).
    \label{eqn:total-potential-marginal-stability}
\end{equation}
%%%%%%%%%

We stress that eq. \eqref{eq:non-divergent-response} is valid for stable \textit{and} unstable systems, 
and provides a smooth transition between them.  
It also provides sensible results in limiting cases. For strongly stable systems, the damping timescale $-1/\gamma_{\bm{k}}$ is short, so the Landau mode term decays rapidly and we 
soon recover the dressed `wake' potential \eqref{eq:response-no-eigenmodes}.
On the other hand, for unstable systems the Landau mode eventually dominates.  Of course, in the unstable case, 
\eqref{eq:non-divergent-response} eventually breaks down once the Landau mode has grown so large that nonlinear effects become important.
%and we recover the 

%On the latter case, the second term in the square brackets

Before moving on, let us simplify \cref{eq:non-divergent-response}
by supposing that (i) the perturbing point mass is either at rest or that the wave number $\bm{k}$ is perpendicular to its velocity, so $\bm{k}\cdot \bm{v}_{\prtcl}=0$, and (ii) the most weakly damped eigenmode has a purely imaginary frequency $z_{\bm{k}}=i\gamma_{\bm{k}}$. Then\footnote{Note that {for $\bm{k}\cdot\bm{v}_{\prtcl}=0$, eq. \eqref{eq:point-mass-potential} implies that $\Phi_{\bm{k}}^{\ext}(t)=\Phi_{\bm{k}}^{\ext}(0)$}.}
\begin{equation}
    \label{eq:response-simplified}
    \Phi_{\bm{k}}^{\tot}(t)
    = \left[
        \frac{1}{\epsilon_{\bm{k}}(0)}
        + \frac{\exp(\gamma_{\bm{k}}t)}
        {z_{\bm{k}}\epsilon_{\bm{k}}'(z_{\bm{k}})}
    \right]
    \Phi_{\bm{k}}^{\ext}(t).
\end{equation}
We will use this expression in the next subsection.

\subsection{Maxwellian stellar system}

\begin{figure}
    \centerline{\input{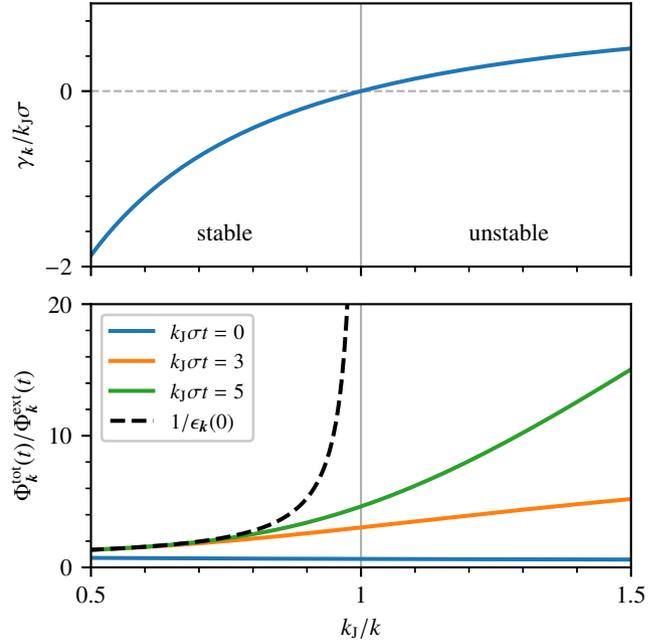}}
    \caption{Top: growth rate of the most weakly damped (or most rapidly growing) mode of a Maxwellian stellar system as a function of dimensionless wavelength
    $k_\mathrm{J}/k$.  Bottom:  we suppose the Maxwellian system is perturbed by a point mass, 
    and plot the ratio of the total potential to the perturber potential, eq. \eqref{eq:maxwellian-response}, for different times $t$.
    The black dashed line shows the naive wake result that ignores the Landau mode contribution (eq. \eqref{eq:response-no-eigenmodes}).}
    \label{fig:maxwellian}
\end{figure}
%\begin{figure}
%    \centerline{\input{figures/maxwellian-2.pgf}}
%\end{figure}
%\begin{figure}
%    \centerline{\input{figures/maxwellian-3.pgf}}
%\end{figure}

Let us suppose that the background DF is Maxwellian, i.e.
\begin{equation}
  f_0(\bm{v}) = \frac{\rho_0}{(2\pi)^{3/2}\sigma^3}
  \exp\left(-\frac{v^2}{2\sigma^2}\right),
\end{equation}
where $\sigma>0$ is the velocity dispersion.
The dispersion function \eqref{eq:permittivity-plus} is given by
\begin{equation}
  \label{eq:pines}
  \epsilon_{\bm{k}}(z)
  = 1 - \frac{\kJ^2}{k^2} W\left(\frac{z}{k\sigma}\right),
\end{equation}
where $\kJ=\sqrt{4\pi G\rho_0}/\sigma$ is the Jeans wavenumber, and
\begin{equation}
  W(\zeta)
  = \frac{1}{\sqrt{2\pi}}\int_{-\infty}^\infty\diff x\,
  \frac{x}{(x - \zeta)^\gobelow}\exp\left(-\frac{x^2}{2}\right)
\end{equation}
is the plasma dispersion function as defined by \citet{Ichimaru1973}.
It is easy to see that $W(0)=1$, and hence that $\epsilon_{\bm{k}}(0)=1-\kJ^2/k^2$.

The plasma dispersion function satisfies the differential equation
$W'(\zeta)=(1/\zeta-\zeta)W(\zeta)-1/\zeta$.
From this together with \cref{eq:pines} it follows that $z_{\bm{k}}\epsilon_{\bm{k}}'(z_{\bm{k}})=z_{\bm{k}}^2/(k\sigma)^2-\epsilon_{\bm{k}}(0)$, where we have used the fact that $\epsilon_{\bm{k}}(z_{\bm{k}})=0$ and hence $W(z_{\bm{k}}/k\sigma)=k^2/\kJ^2$.
For a Maxwellian, \cref{eq:response-simplified} thus reads
\begin{equation}
    \label{eq:maxwellian-response}
    \frac{\Phi_{\bm{k}}^{\tot}(t)}{\Phi_{\bm{k}}^{\ext}(t)}
    =     \frac{1}{1 - \kJ^2/k^2}
    - \frac{\exp(-iz_{\bm{k}}t)}{1 - \kJ^2/k^2 - z_{\bm{k}}^2/(k\sigma)^2}.
\end{equation}

The small argument expansion of the plasma dispersion function is $W(\zeta)=1+i\sqrt{\pi/2}\,\zeta+O(\zeta^2)$.
From this it follows that near marginal stability, the Landau mode frequency is $z_{\bm{k}}=i\gamma_{\bm{k}}$ with
\begin{equation}
  \gamma_{\bm{k}} = \sqrt{\frac{8}{\pi}}(\kJ - k)\sigma.
  \label{eqn:growth-rate-near-marginal-stability}
\end{equation}
Obviously the system is stable (unstable) for $k>k_\mathrm{J}$ ($k<k_\mathrm{J}$).
Substituting \eqref{eqn:growth-rate-near-marginal-stability} into \cref{eq:maxwellian-response} and taking the limit {of marginal stability},
$k\to\kJ$, yields
\begin{equation}
    \lim_{k\to\kJ}
     \frac{\Phi_{\bm{k}}^{\tot}(t)}{\Phi_{\bm{k}}^{\ext}(t)}
    = \frac{2}{\pi} + \sqrt{\frac{2}{\pi}}\,k\sigma t.
\end{equation}

For wavenumbers that are significantly different from $\kJ$ we can compute
the growth rate $\gamma_{\bm{k}}$ numerically using a root-finding algorithm, and hence calculate $\Phi_{\bm{k}}^\mathrm{tot}(t)$
using eq. \eqref{eq:maxwellian-response}. 
We show the result of this calculation in \cref{fig:maxwellian}.  The top panel shows the growth rate $\gamma_{\bm{k}}$ as a function of scale $k_\mathrm{J}/k$; as expected, the $k$-space divides into stable ($k_\mathrm{J}/k < 1$) and unstable ($k_\mathrm{J}/k > 1$) regions.
In the bottom panel we plot with different coloured lines the ratio of the total potential to external potential, $\Phi_{\bm{k}}^\mathrm{tot}(t)/\Phi_{\bm{k}}^\mathrm{ext}(t)$, computed using eq. \eqref{eq:maxwellian-response} at different times $t$.
With a black dashed line we show the (time-independent) part of 
\eqref{eq:maxwellian-response} that arises by considering 
only the first term (the wake contribution) and ignoring the second term (the Landau mode).
In other words, the black dashed line is the naive 
result that one would find {by simply applying} \eqref{eq:response-no-eigenmodes} in the weakly stable regime {(as is often done in Balescu-Lenard calculations, e.g. \citealt{fouvry2019secular})}.
We see clearly that {this contribution is divergent for marginally stable systems ($k_\mathrm{J}/k \to 1$), and severely overestimates the linear response for weakly stable systems (say $k_\mathrm{J}/k \approx 0.95$)
at early times}. 
This emphasises the fact that even for stable systems, the form 
\eqref{eq:response-no-eigenmodes} of the external potential {is only valid for times $t \gg -1/\gamma_{\bm{k}}$, which becomes infinite as marginal stability is approached.}

Note that we do not recover the exact initial condition
$\Phi_{\bm{k}}^\mathrm{tot}(0) = \Phi^\mathrm{ext}_{\bm{k}}(0) $ for all $k$ in \cref{fig:maxwellian} (see the blue line in the bottom panel).  Instead, $\Phi_{\bm{k}}^\mathrm{tot}(0)/\Phi^\mathrm{ext}_{\bm{k}}(0) $ equals unity only for $k_\mathrm{J}/k=0$, and then decreases slowly with increasing $k_\mathrm{J}/k$.
The reason for this is that when we wrote down eq. \eqref{eq:non-divergent-response},
we accounted only for the system's most weakly damped Landau mode 
and ignored the contributions from  any other modes.
Had we included all the modes, we would find $\Phi_{\bm{k}}^\mathrm{tot}(0)/\Phi^\mathrm{ext}_{\bm{k}}(0) = 1$ for all $k$.
We have checked this claim
explicitly for a Lorentzian stellar system (eq. \eqref{eqn:kappa} with $\kappa = 1$), 
for which all modes can be calculated analytically.
Apart 
from this minor technical detail, the linear response behavior of
the Lorentzian stellar system {does} not differ quantitatively that of the Maxwellian, so we do not report the details here. 

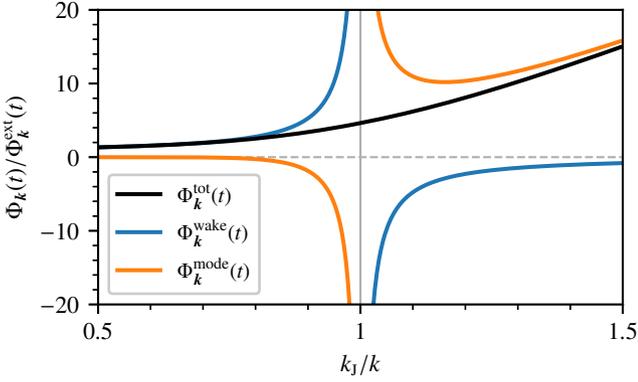
\begin{figure}
    \centerline{\input{figures/divergences.pgf}}
    \caption{Plot of the separate wake (blue) and mode (orange) contributions to $\Phi_{\bm{k}}^\mathrm{tot}(t)/\Phi_{\bm{k}}^\mathrm{ext}(t)$ (see equations \eqref{eqn:wake_and_mode})
    at fixed time $\kJ\sigma t=5$. The black line shows the sum of the two contributions.}
    \label{fig:divergences}
\end{figure}

Finally, in \cref{fig:divergences} we fix $k_\mathrm{J} \sigma t = 5$,
and show explicitly the contributions from the two individual terms in
\cref{eq:maxwellian-response}, namely
\begin{equation}
    \Phi_{\bm{k}}^{\mathrm{wake}}(t) = \frac{\Phi_{\bm{k}}^{\ext}(t)}{1 - \kJ^2/k^2}
    \ \ \,\text{and}\,\ \ 
    \Phi_{\bm{k}}^{\mathrm{mode}}(t)
    = -\frac{\exp(-iz_{\bm{k}}t)\,\Phi_{\bm{k}}^{\ext}(t)}{1 - \kJ^2/k^2 - z_{\bm{k}}^2/(k\sigma)^2},
    \label{eqn:wake_and_mode}
\end{equation}
respectively, as well as their sum.
%(Note that the `wake' contribution contains the potential of the perturber, which would need to be subtracted if we wanted to isolate the induced wake arising from disturbed particles only.)
We see that in the strongly stable regime $k_\mathrm{J}/k \ll 1$, the mode contribution is negligible, 
and the wake contribution is just equal to the perturber potential ($\Phi_{\bm{k}}^\mathrm{tot}(0)/\Phi^\mathrm{ext}_{\bm{k}}(0) \to  1$).
This makes sense since at small scales, self-gravity is unimportant, so $\epsilon_{\bm{k}}(z) \approx 1$ for all $z$.
On the other hand, in the strongly unstable regime $k_\mathrm{J}/k \gg 1$, the exponentially-growing 
mode contribution dominates the total potential.  Both the `wake' and `mode' contributions individually diverge as $k_\mathrm{J}/k$ approaches unity from either side, but their sum is perfectly well-defined for all $k$.  None of our calculations exhibit any sharp feature {at marginal stability}.

%%%%%%%%%%%%%%%%%%%%%%%%%%%%%%%%%%%%%%%%%%%%%%%%%%%%%%%%%%%%%%%%%%%%%%%%
%%%%%%%%%%%%%%%%%%%%%%%%%%%%%%%%%%%%%%%%%%%%%%%%%%%%%%%%%%%%%%%%%%%%%%%%
%%%%%%%%%%%%%%%%%%%%%%%%%%%%%%%%%%%%%%%%%%%%%%%%%%%%%%%%%%%%%%%%%%%%%%%%

\section{Summary}
\label{sec:Summary}

The self-consistent response of a gravitating system to a weak perturbation 
is a fundamental problem of stellar dynamics.
The most basic mathematical tool with which this problem may be attacked 
is linear response theory \citep{palmer1994stability,Nelson1999-in,Binney2008-ou}.
However, several previous calculations based on the dressed `wakes' that perturbations induce in stellar systems
have suggested, implicitly or explicitly, that the linear 
response potential actually diverges if the system in question happens to {approach 
marginally stability} (see the Introduction for examples).
Such an obviously unphysical result might lead one to conclude that, for {very weakly} stable stellar systems at least, linear theory ought to be abandoned.  More optimistically, one might hope that the divergence can be cured (e.g. by nonlinear effects), but that wakes are nevertheless very efficient means to amplify gravitational perturbations, and hence drive rapid secular evolution in galaxies \citep{Fouvry2015-nk,chavanis2023secular}.

In this paper we demonstrated that {for finite times}, the linear response theory is in fact not divergent at all, regardless of a stellar system's stability properties.  To do this, 
we calculated
the linear gravitational potential response of a homogeneous stellar system to a point mass perturber.
The result
naturally decomposes into a `wake'-like part associated with polarization (`dressing') of the perturber potential,
and an exponentially growing or decaying response which is associated with the excitation of the system's Landau modes.
We showed analytically that as the system approaches marginal stability (i.e. as the imaginary part of its most weakly damped Landau mode frequency approaches zero), the contributions from the dressed wake and the mode both individually diverge, but their sum, which is the only physically relevant
quantity, does not.  
This fact was already understood many years ago in plasma physics (e.g. \citealt{Rogister1968-tb,hatori1969nonstationary,oberman1970advanced}), 
but does not seem to have been appreciated in stellar dynamics until now.
As an example, we considered the linear response of a Maxwellian stellar system to an orbiting point mass.
{We} illustrated explicitly the cancellation of wake and mode divergences in the vicinity of marginal stability {at any finite time}, and showed that the 
total potential response behaves smoothly at all scales $k$.  {The key point is that the divergent result \eqref{eq:response-no-eigenmodes} employed in previous studies is only valid on timescales long compared to the decay time of all Landau modes, and for marginally stable systems no such timescale exists.}

While we have focused exclusively on homogeneous stellar systems perturbed by a single point mass, one may straightforwardly extend our results to inhomogeneous stellar systems, to multiple perturbers, 
and to perturbations that are internally generated rather than externally imposed. 
Such calculations will form part of our future work; here we simply state the generic result that 
the {finite-time} linear response of stellar systems never diverges as a result of the system approaching marginal stability.
This implies that
\textit{the total, self-gravitating response of galaxies may be much weaker than would be naively calculated based on dressed wakes alone}.
This is a good thing from the point of view of the validity of linear response theory, which would of course be rendered invalid if the
the fluctuations in weakly stable systems really were extremely (i.e. nonlinearly) large.
On the other hand, it also suggests that wake-driven secular evolution of galaxies may be much less efficient than previously thought, 
and the rate of this evolution may be severely overestimated by kinetic schemes that explicitly ignore the direct contribution from Landau modes, like Balescu-Lenard theory \citep{heyvaerts2010balescu,chavanis2012kinetic,Hamilton2021-qe}. In upcoming work, we will present a unified kinetic theory that accounts for both types of response self-consistently \citep{Rogister1968-tb,hatori1969nonstationary,oberman1970advanced}.

In conclusion, any previous calculation of the evolution of a stellar system
that accounted for strongly amplified wakes, but ignored the explicit contribution from Landau modes,
may need to be revised.
%In upcoming work we will present a unified kinetic theory that accounts for both types of response self-consistently.

%%%%%%%%%%%%%%
%%%%%%%%%%%%%%
%%%%%%%%%%%%%%
%%%%%%%%%%%%%%
%%%%%%%%%%%%%%
%%%%%%%%%%%%%%
%%%%%%%%%%%%%%
%%%%%%%%%%%%%%

\section*{Acknowledgements}

{We thank James Binney, John Magorrian, Scott Tremaine, Lev Arzamasskiy and the anonymous referee for several useful comments.}
This work was supported by a grant from the Simons Foundation (816048, CH).

\section*{Data availability}
No new data were generated or analysed in support of this research.

%%%%%%%%%%%%%%%%%%%%%%%%%%%%%%%%%%%%%%%%%%%%%%%%%%



\bibliographystyle{mnras}
\bibliography{Bibliography} % if your bibtex file is called example.bib

%\appendix

%%%%%%%%%%%%%%%%%%%%%%%%%%%%%%%%%%%%%%%%%%%%%%%%%%%%%%%%%%%%

% Don't change these lines
\bsp	% typesetting comment
\label{lastpage}

\end{document}

%% file: figures/divergences.pgf
%% Creator: Matplotlib, PGF backend
%%
%% To include the figure in your LaTeX document, write
%%   \input{<filename>.pgf}
%%
%% Make sure the required packages are loaded in your preamble
%%   \usepackage{pgf}
%%
%% Also ensure that all the required font packages are loaded; for instance,
%% the lmodern package is sometimes necessary when using math font.
%%   \usepackage{lmodern}
%%
%% Figures using additional raster images can only be included by \input if
%% they are in the same directory as the main LaTeX file. For loading figures
%% from other directories you can use the `import` package
%% and then include the figures with
%%   \import{<path to file>}{<filename>.pgf}
%%
%% Matplotlib used the following preamble
%%   
%%           \usepackage[T1]{fontenc}
%%           \usepackage[utf8]{inputenc}
%%           \usepackage{amsmath}
%%           \usepackage{txfonts}
%%           \usepackage{bm}
%%       
%%   \usepackage{fontspec}
%%   \makeatletter\@ifpackageloaded{underscore}{}{\usepackage[strings]{underscore}}\makeatother
%%
\begingroup%
\makeatletter%
\begin{pgfpicture}%
\pgfpathrectangle{\pgfpointorigin}{\pgfqpoint{3.376228in}{2.025737in}}%
\pgfusepath{use as bounding box, clip}%
\begin{pgfscope}%
\pgfsetbuttcap%
\pgfsetmiterjoin%
\definecolor{currentfill}{rgb}{1.000000,1.000000,1.000000}%
\pgfsetfillcolor{currentfill}%
\pgfsetlinewidth{0.000000pt}%
\definecolor{currentstroke}{rgb}{1.000000,1.000000,1.000000}%
\pgfsetstrokecolor{currentstroke}%
\pgfsetdash{}{0pt}%
\pgfpathmoveto{\pgfqpoint{0.000000in}{0.000000in}}%
\pgfpathlineto{\pgfqpoint{3.376228in}{0.000000in}}%
\pgfpathlineto{\pgfqpoint{3.376228in}{2.025737in}}%
\pgfpathlineto{\pgfqpoint{0.000000in}{2.025737in}}%
\pgfpathlineto{\pgfqpoint{0.000000in}{0.000000in}}%
\pgfpathclose%
\pgfusepath{fill}%
\end{pgfscope}%
\begin{pgfscope}%
\pgfsetbuttcap%
\pgfsetmiterjoin%
\definecolor{currentfill}{rgb}{1.000000,1.000000,1.000000}%
\pgfsetfillcolor{currentfill}%
\pgfsetlinewidth{0.000000pt}%
\definecolor{currentstroke}{rgb}{0.000000,0.000000,0.000000}%
\pgfsetstrokecolor{currentstroke}%
\pgfsetstrokeopacity{0.000000}%
\pgfsetdash{}{0pt}%
\pgfpathmoveto{\pgfqpoint{0.534489in}{0.416447in}}%
\pgfpathlineto{\pgfqpoint{3.252433in}{0.416447in}}%
\pgfpathlineto{\pgfqpoint{3.252433in}{1.940692in}}%
\pgfpathlineto{\pgfqpoint{0.534489in}{1.940692in}}%
\pgfpathlineto{\pgfqpoint{0.534489in}{0.416447in}}%
\pgfpathclose%
\pgfusepath{fill}%
\end{pgfscope}%
\begin{pgfscope}%
\pgfsetbuttcap%
\pgfsetroundjoin%
\definecolor{currentfill}{rgb}{0.000000,0.000000,0.000000}%
\pgfsetfillcolor{currentfill}%
\pgfsetlinewidth{0.803000pt}%
\definecolor{currentstroke}{rgb}{0.000000,0.000000,0.000000}%
\pgfsetstrokecolor{currentstroke}%
\pgfsetdash{}{0pt}%
\pgfsys@defobject{currentmarker}{\pgfqpoint{0.000000in}{-0.048611in}}{\pgfqpoint{0.000000in}{0.000000in}}{%
\pgfpathmoveto{\pgfqpoint{0.000000in}{0.000000in}}%
\pgfpathlineto{\pgfqpoint{0.000000in}{-0.048611in}}%
\pgfusepath{stroke,fill}%
}%
\begin{pgfscope}%
\pgfsys@transformshift{0.534489in}{0.416447in}%
\pgfsys@useobject{currentmarker}{}%
\end{pgfscope}%
\end{pgfscope}%
\begin{pgfscope}%
\definecolor{textcolor}{rgb}{0.000000,0.000000,0.000000}%
\pgfsetstrokecolor{textcolor}%
\pgfsetfillcolor{textcolor}%
\pgftext[x=0.534489in,y=0.319225in,,top]{\color{textcolor}\rmfamily\fontsize{9.000000}{10.800000}\selectfont 0.5}%
\end{pgfscope}%
\begin{pgfscope}%
\pgfsetbuttcap%
\pgfsetroundjoin%
\definecolor{currentfill}{rgb}{0.000000,0.000000,0.000000}%
\pgfsetfillcolor{currentfill}%
\pgfsetlinewidth{0.803000pt}%
\definecolor{currentstroke}{rgb}{0.000000,0.000000,0.000000}%
\pgfsetstrokecolor{currentstroke}%
\pgfsetdash{}{0pt}%
\pgfsys@defobject{currentmarker}{\pgfqpoint{0.000000in}{-0.048611in}}{\pgfqpoint{0.000000in}{0.000000in}}{%
\pgfpathmoveto{\pgfqpoint{0.000000in}{0.000000in}}%
\pgfpathlineto{\pgfqpoint{0.000000in}{-0.048611in}}%
\pgfusepath{stroke,fill}%
}%
\begin{pgfscope}%
\pgfsys@transformshift{1.893461in}{0.416447in}%
\pgfsys@useobject{currentmarker}{}%
\end{pgfscope}%
\end{pgfscope}%
\begin{pgfscope}%
\definecolor{textcolor}{rgb}{0.000000,0.000000,0.000000}%
\pgfsetstrokecolor{textcolor}%
\pgfsetfillcolor{textcolor}%
\pgftext[x=1.893461in,y=0.319225in,,top]{\color{textcolor}\rmfamily\fontsize{9.000000}{10.800000}\selectfont 1}%
\end{pgfscope}%
\begin{pgfscope}%
\pgfsetbuttcap%
\pgfsetroundjoin%
\definecolor{currentfill}{rgb}{0.000000,0.000000,0.000000}%
\pgfsetfillcolor{currentfill}%
\pgfsetlinewidth{0.803000pt}%
\definecolor{currentstroke}{rgb}{0.000000,0.000000,0.000000}%
\pgfsetstrokecolor{currentstroke}%
\pgfsetdash{}{0pt}%
\pgfsys@defobject{currentmarker}{\pgfqpoint{0.000000in}{-0.048611in}}{\pgfqpoint{0.000000in}{0.000000in}}{%
\pgfpathmoveto{\pgfqpoint{0.000000in}{0.000000in}}%
\pgfpathlineto{\pgfqpoint{0.000000in}{-0.048611in}}%
\pgfusepath{stroke,fill}%
}%
\begin{pgfscope}%
\pgfsys@transformshift{3.252433in}{0.416447in}%
\pgfsys@useobject{currentmarker}{}%
\end{pgfscope}%
\end{pgfscope}%
\begin{pgfscope}%
\definecolor{textcolor}{rgb}{0.000000,0.000000,0.000000}%
\pgfsetstrokecolor{textcolor}%
\pgfsetfillcolor{textcolor}%
\pgftext[x=3.252433in,y=0.319225in,,top]{\color{textcolor}\rmfamily\fontsize{9.000000}{10.800000}\selectfont 1.5}%
\end{pgfscope}%
\begin{pgfscope}%
\pgfsetbuttcap%
\pgfsetroundjoin%
\definecolor{currentfill}{rgb}{0.000000,0.000000,0.000000}%
\pgfsetfillcolor{currentfill}%
\pgfsetlinewidth{0.602250pt}%
\definecolor{currentstroke}{rgb}{0.000000,0.000000,0.000000}%
\pgfsetstrokecolor{currentstroke}%
\pgfsetdash{}{0pt}%
\pgfsys@defobject{currentmarker}{\pgfqpoint{0.000000in}{-0.027778in}}{\pgfqpoint{0.000000in}{0.000000in}}{%
\pgfpathmoveto{\pgfqpoint{0.000000in}{0.000000in}}%
\pgfpathlineto{\pgfqpoint{0.000000in}{-0.027778in}}%
\pgfusepath{stroke,fill}%
}%
\begin{pgfscope}%
\pgfsys@transformshift{0.806283in}{0.416447in}%
\pgfsys@useobject{currentmarker}{}%
\end{pgfscope}%
\end{pgfscope}%
\begin{pgfscope}%
\pgfsetbuttcap%
\pgfsetroundjoin%
\definecolor{currentfill}{rgb}{0.000000,0.000000,0.000000}%
\pgfsetfillcolor{currentfill}%
\pgfsetlinewidth{0.602250pt}%
\definecolor{currentstroke}{rgb}{0.000000,0.000000,0.000000}%
\pgfsetstrokecolor{currentstroke}%
\pgfsetdash{}{0pt}%
\pgfsys@defobject{currentmarker}{\pgfqpoint{0.000000in}{-0.027778in}}{\pgfqpoint{0.000000in}{0.000000in}}{%
\pgfpathmoveto{\pgfqpoint{0.000000in}{0.000000in}}%
\pgfpathlineto{\pgfqpoint{0.000000in}{-0.027778in}}%
\pgfusepath{stroke,fill}%
}%
\begin{pgfscope}%
\pgfsys@transformshift{1.078077in}{0.416447in}%
\pgfsys@useobject{currentmarker}{}%
\end{pgfscope}%
\end{pgfscope}%
\begin{pgfscope}%
\pgfsetbuttcap%
\pgfsetroundjoin%
\definecolor{currentfill}{rgb}{0.000000,0.000000,0.000000}%
\pgfsetfillcolor{currentfill}%
\pgfsetlinewidth{0.602250pt}%
\definecolor{currentstroke}{rgb}{0.000000,0.000000,0.000000}%
\pgfsetstrokecolor{currentstroke}%
\pgfsetdash{}{0pt}%
\pgfsys@defobject{currentmarker}{\pgfqpoint{0.000000in}{-0.027778in}}{\pgfqpoint{0.000000in}{0.000000in}}{%
\pgfpathmoveto{\pgfqpoint{0.000000in}{0.000000in}}%
\pgfpathlineto{\pgfqpoint{0.000000in}{-0.027778in}}%
\pgfusepath{stroke,fill}%
}%
\begin{pgfscope}%
\pgfsys@transformshift{1.349872in}{0.416447in}%
\pgfsys@useobject{currentmarker}{}%
\end{pgfscope}%
\end{pgfscope}%
\begin{pgfscope}%
\pgfsetbuttcap%
\pgfsetroundjoin%
\definecolor{currentfill}{rgb}{0.000000,0.000000,0.000000}%
\pgfsetfillcolor{currentfill}%
\pgfsetlinewidth{0.602250pt}%
\definecolor{currentstroke}{rgb}{0.000000,0.000000,0.000000}%
\pgfsetstrokecolor{currentstroke}%
\pgfsetdash{}{0pt}%
\pgfsys@defobject{currentmarker}{\pgfqpoint{0.000000in}{-0.027778in}}{\pgfqpoint{0.000000in}{0.000000in}}{%
\pgfpathmoveto{\pgfqpoint{0.000000in}{0.000000in}}%
\pgfpathlineto{\pgfqpoint{0.000000in}{-0.027778in}}%
\pgfusepath{stroke,fill}%
}%
\begin{pgfscope}%
\pgfsys@transformshift{1.621666in}{0.416447in}%
\pgfsys@useobject{currentmarker}{}%
\end{pgfscope}%
\end{pgfscope}%
\begin{pgfscope}%
\pgfsetbuttcap%
\pgfsetroundjoin%
\definecolor{currentfill}{rgb}{0.000000,0.000000,0.000000}%
\pgfsetfillcolor{currentfill}%
\pgfsetlinewidth{0.602250pt}%
\definecolor{currentstroke}{rgb}{0.000000,0.000000,0.000000}%
\pgfsetstrokecolor{currentstroke}%
\pgfsetdash{}{0pt}%
\pgfsys@defobject{currentmarker}{\pgfqpoint{0.000000in}{-0.027778in}}{\pgfqpoint{0.000000in}{0.000000in}}{%
\pgfpathmoveto{\pgfqpoint{0.000000in}{0.000000in}}%
\pgfpathlineto{\pgfqpoint{0.000000in}{-0.027778in}}%
\pgfusepath{stroke,fill}%
}%
\begin{pgfscope}%
\pgfsys@transformshift{2.165255in}{0.416447in}%
\pgfsys@useobject{currentmarker}{}%
\end{pgfscope}%
\end{pgfscope}%
\begin{pgfscope}%
\pgfsetbuttcap%
\pgfsetroundjoin%
\definecolor{currentfill}{rgb}{0.000000,0.000000,0.000000}%
\pgfsetfillcolor{currentfill}%
\pgfsetlinewidth{0.602250pt}%
\definecolor{currentstroke}{rgb}{0.000000,0.000000,0.000000}%
\pgfsetstrokecolor{currentstroke}%
\pgfsetdash{}{0pt}%
\pgfsys@defobject{currentmarker}{\pgfqpoint{0.000000in}{-0.027778in}}{\pgfqpoint{0.000000in}{0.000000in}}{%
\pgfpathmoveto{\pgfqpoint{0.000000in}{0.000000in}}%
\pgfpathlineto{\pgfqpoint{0.000000in}{-0.027778in}}%
\pgfusepath{stroke,fill}%
}%
\begin{pgfscope}%
\pgfsys@transformshift{2.437050in}{0.416447in}%
\pgfsys@useobject{currentmarker}{}%
\end{pgfscope}%
\end{pgfscope}%
\begin{pgfscope}%
\pgfsetbuttcap%
\pgfsetroundjoin%
\definecolor{currentfill}{rgb}{0.000000,0.000000,0.000000}%
\pgfsetfillcolor{currentfill}%
\pgfsetlinewidth{0.602250pt}%
\definecolor{currentstroke}{rgb}{0.000000,0.000000,0.000000}%
\pgfsetstrokecolor{currentstroke}%
\pgfsetdash{}{0pt}%
\pgfsys@defobject{currentmarker}{\pgfqpoint{0.000000in}{-0.027778in}}{\pgfqpoint{0.000000in}{0.000000in}}{%
\pgfpathmoveto{\pgfqpoint{0.000000in}{0.000000in}}%
\pgfpathlineto{\pgfqpoint{0.000000in}{-0.027778in}}%
\pgfusepath{stroke,fill}%
}%
\begin{pgfscope}%
\pgfsys@transformshift{2.708844in}{0.416447in}%
\pgfsys@useobject{currentmarker}{}%
\end{pgfscope}%
\end{pgfscope}%
\begin{pgfscope}%
\pgfsetbuttcap%
\pgfsetroundjoin%
\definecolor{currentfill}{rgb}{0.000000,0.000000,0.000000}%
\pgfsetfillcolor{currentfill}%
\pgfsetlinewidth{0.602250pt}%
\definecolor{currentstroke}{rgb}{0.000000,0.000000,0.000000}%
\pgfsetstrokecolor{currentstroke}%
\pgfsetdash{}{0pt}%
\pgfsys@defobject{currentmarker}{\pgfqpoint{0.000000in}{-0.027778in}}{\pgfqpoint{0.000000in}{0.000000in}}{%
\pgfpathmoveto{\pgfqpoint{0.000000in}{0.000000in}}%
\pgfpathlineto{\pgfqpoint{0.000000in}{-0.027778in}}%
\pgfusepath{stroke,fill}%
}%
\begin{pgfscope}%
\pgfsys@transformshift{2.980639in}{0.416447in}%
\pgfsys@useobject{currentmarker}{}%
\end{pgfscope}%
\end{pgfscope}%
\begin{pgfscope}%
\definecolor{textcolor}{rgb}{0.000000,0.000000,0.000000}%
\pgfsetstrokecolor{textcolor}%
\pgfsetfillcolor{textcolor}%
\pgftext[x=1.893461in,y=0.152670in,,top]{\color{textcolor}\rmfamily\fontsize{9.000000}{10.800000}\selectfont \(\displaystyle k_{\mathrm{J}}/k\)}%
\end{pgfscope}%
\begin{pgfscope}%
\pgfsetbuttcap%
\pgfsetroundjoin%
\definecolor{currentfill}{rgb}{0.000000,0.000000,0.000000}%
\pgfsetfillcolor{currentfill}%
\pgfsetlinewidth{0.803000pt}%
\definecolor{currentstroke}{rgb}{0.000000,0.000000,0.000000}%
\pgfsetstrokecolor{currentstroke}%
\pgfsetdash{}{0pt}%
\pgfsys@defobject{currentmarker}{\pgfqpoint{-0.048611in}{0.000000in}}{\pgfqpoint{-0.000000in}{0.000000in}}{%
\pgfpathmoveto{\pgfqpoint{-0.000000in}{0.000000in}}%
\pgfpathlineto{\pgfqpoint{-0.048611in}{0.000000in}}%
\pgfusepath{stroke,fill}%
}%
\begin{pgfscope}%
\pgfsys@transformshift{0.534489in}{0.416447in}%
\pgfsys@useobject{currentmarker}{}%
\end{pgfscope}%
\end{pgfscope}%
\begin{pgfscope}%
\definecolor{textcolor}{rgb}{0.000000,0.000000,0.000000}%
\pgfsetstrokecolor{textcolor}%
\pgfsetfillcolor{textcolor}%
\pgftext[x=0.232767in, y=0.373072in, left, base]{\color{textcolor}\rmfamily\fontsize{9.000000}{10.800000}\selectfont \(\displaystyle {\ensuremath{-}20}\)}%
\end{pgfscope}%
\begin{pgfscope}%
\pgfsetbuttcap%
\pgfsetroundjoin%
\definecolor{currentfill}{rgb}{0.000000,0.000000,0.000000}%
\pgfsetfillcolor{currentfill}%
\pgfsetlinewidth{0.803000pt}%
\definecolor{currentstroke}{rgb}{0.000000,0.000000,0.000000}%
\pgfsetstrokecolor{currentstroke}%
\pgfsetdash{}{0pt}%
\pgfsys@defobject{currentmarker}{\pgfqpoint{-0.048611in}{0.000000in}}{\pgfqpoint{-0.000000in}{0.000000in}}{%
\pgfpathmoveto{\pgfqpoint{-0.000000in}{0.000000in}}%
\pgfpathlineto{\pgfqpoint{-0.048611in}{0.000000in}}%
\pgfusepath{stroke,fill}%
}%
\begin{pgfscope}%
\pgfsys@transformshift{0.534489in}{0.797509in}%
\pgfsys@useobject{currentmarker}{}%
\end{pgfscope}%
\end{pgfscope}%
\begin{pgfscope}%
\definecolor{textcolor}{rgb}{0.000000,0.000000,0.000000}%
\pgfsetstrokecolor{textcolor}%
\pgfsetfillcolor{textcolor}%
\pgftext[x=0.232767in, y=0.754134in, left, base]{\color{textcolor}\rmfamily\fontsize{9.000000}{10.800000}\selectfont \(\displaystyle {\ensuremath{-}10}\)}%
\end{pgfscope}%
\begin{pgfscope}%
\pgfsetbuttcap%
\pgfsetroundjoin%
\definecolor{currentfill}{rgb}{0.000000,0.000000,0.000000}%
\pgfsetfillcolor{currentfill}%
\pgfsetlinewidth{0.803000pt}%
\definecolor{currentstroke}{rgb}{0.000000,0.000000,0.000000}%
\pgfsetstrokecolor{currentstroke}%
\pgfsetdash{}{0pt}%
\pgfsys@defobject{currentmarker}{\pgfqpoint{-0.048611in}{0.000000in}}{\pgfqpoint{-0.000000in}{0.000000in}}{%
\pgfpathmoveto{\pgfqpoint{-0.000000in}{0.000000in}}%
\pgfpathlineto{\pgfqpoint{-0.048611in}{0.000000in}}%
\pgfusepath{stroke,fill}%
}%
\begin{pgfscope}%
\pgfsys@transformshift{0.534489in}{1.178570in}%
\pgfsys@useobject{currentmarker}{}%
\end{pgfscope}%
\end{pgfscope}%
\begin{pgfscope}%
\definecolor{textcolor}{rgb}{0.000000,0.000000,0.000000}%
\pgfsetstrokecolor{textcolor}%
\pgfsetfillcolor{textcolor}%
\pgftext[x=0.374766in, y=1.135195in, left, base]{\color{textcolor}\rmfamily\fontsize{9.000000}{10.800000}\selectfont \(\displaystyle {0}\)}%
\end{pgfscope}%
\begin{pgfscope}%
\pgfsetbuttcap%
\pgfsetroundjoin%
\definecolor{currentfill}{rgb}{0.000000,0.000000,0.000000}%
\pgfsetfillcolor{currentfill}%
\pgfsetlinewidth{0.803000pt}%
\definecolor{currentstroke}{rgb}{0.000000,0.000000,0.000000}%
\pgfsetstrokecolor{currentstroke}%
\pgfsetdash{}{0pt}%
\pgfsys@defobject{currentmarker}{\pgfqpoint{-0.048611in}{0.000000in}}{\pgfqpoint{-0.000000in}{0.000000in}}{%
\pgfpathmoveto{\pgfqpoint{-0.000000in}{0.000000in}}%
\pgfpathlineto{\pgfqpoint{-0.048611in}{0.000000in}}%
\pgfusepath{stroke,fill}%
}%
\begin{pgfscope}%
\pgfsys@transformshift{0.534489in}{1.559631in}%
\pgfsys@useobject{currentmarker}{}%
\end{pgfscope}%
\end{pgfscope}%
\begin{pgfscope}%
\definecolor{textcolor}{rgb}{0.000000,0.000000,0.000000}%
\pgfsetstrokecolor{textcolor}%
\pgfsetfillcolor{textcolor}%
\pgftext[x=0.312266in, y=1.516256in, left, base]{\color{textcolor}\rmfamily\fontsize{9.000000}{10.800000}\selectfont \(\displaystyle {10}\)}%
\end{pgfscope}%
\begin{pgfscope}%
\pgfsetbuttcap%
\pgfsetroundjoin%
\definecolor{currentfill}{rgb}{0.000000,0.000000,0.000000}%
\pgfsetfillcolor{currentfill}%
\pgfsetlinewidth{0.803000pt}%
\definecolor{currentstroke}{rgb}{0.000000,0.000000,0.000000}%
\pgfsetstrokecolor{currentstroke}%
\pgfsetdash{}{0pt}%
\pgfsys@defobject{currentmarker}{\pgfqpoint{-0.048611in}{0.000000in}}{\pgfqpoint{-0.000000in}{0.000000in}}{%
\pgfpathmoveto{\pgfqpoint{-0.000000in}{0.000000in}}%
\pgfpathlineto{\pgfqpoint{-0.048611in}{0.000000in}}%
\pgfusepath{stroke,fill}%
}%
\begin{pgfscope}%
\pgfsys@transformshift{0.534489in}{1.940692in}%
\pgfsys@useobject{currentmarker}{}%
\end{pgfscope}%
\end{pgfscope}%
\begin{pgfscope}%
\definecolor{textcolor}{rgb}{0.000000,0.000000,0.000000}%
\pgfsetstrokecolor{textcolor}%
\pgfsetfillcolor{textcolor}%
\pgftext[x=0.312266in, y=1.897317in, left, base]{\color{textcolor}\rmfamily\fontsize{9.000000}{10.800000}\selectfont \(\displaystyle {20}\)}%
\end{pgfscope}%
\begin{pgfscope}%
\pgfsetbuttcap%
\pgfsetroundjoin%
\definecolor{currentfill}{rgb}{0.000000,0.000000,0.000000}%
\pgfsetfillcolor{currentfill}%
\pgfsetlinewidth{0.602250pt}%
\definecolor{currentstroke}{rgb}{0.000000,0.000000,0.000000}%
\pgfsetstrokecolor{currentstroke}%
\pgfsetdash{}{0pt}%
\pgfsys@defobject{currentmarker}{\pgfqpoint{-0.027778in}{0.000000in}}{\pgfqpoint{-0.000000in}{0.000000in}}{%
\pgfpathmoveto{\pgfqpoint{-0.000000in}{0.000000in}}%
\pgfpathlineto{\pgfqpoint{-0.027778in}{0.000000in}}%
\pgfusepath{stroke,fill}%
}%
\begin{pgfscope}%
\pgfsys@transformshift{0.534489in}{0.492660in}%
\pgfsys@useobject{currentmarker}{}%
\end{pgfscope}%
\end{pgfscope}%
\begin{pgfscope}%
\pgfsetbuttcap%
\pgfsetroundjoin%
\definecolor{currentfill}{rgb}{0.000000,0.000000,0.000000}%
\pgfsetfillcolor{currentfill}%
\pgfsetlinewidth{0.602250pt}%
\definecolor{currentstroke}{rgb}{0.000000,0.000000,0.000000}%
\pgfsetstrokecolor{currentstroke}%
\pgfsetdash{}{0pt}%
\pgfsys@defobject{currentmarker}{\pgfqpoint{-0.027778in}{0.000000in}}{\pgfqpoint{-0.000000in}{0.000000in}}{%
\pgfpathmoveto{\pgfqpoint{-0.000000in}{0.000000in}}%
\pgfpathlineto{\pgfqpoint{-0.027778in}{0.000000in}}%
\pgfusepath{stroke,fill}%
}%
\begin{pgfscope}%
\pgfsys@transformshift{0.534489in}{0.568872in}%
\pgfsys@useobject{currentmarker}{}%
\end{pgfscope}%
\end{pgfscope}%
\begin{pgfscope}%
\pgfsetbuttcap%
\pgfsetroundjoin%
\definecolor{currentfill}{rgb}{0.000000,0.000000,0.000000}%
\pgfsetfillcolor{currentfill}%
\pgfsetlinewidth{0.602250pt}%
\definecolor{currentstroke}{rgb}{0.000000,0.000000,0.000000}%
\pgfsetstrokecolor{currentstroke}%
\pgfsetdash{}{0pt}%
\pgfsys@defobject{currentmarker}{\pgfqpoint{-0.027778in}{0.000000in}}{\pgfqpoint{-0.000000in}{0.000000in}}{%
\pgfpathmoveto{\pgfqpoint{-0.000000in}{0.000000in}}%
\pgfpathlineto{\pgfqpoint{-0.027778in}{0.000000in}}%
\pgfusepath{stroke,fill}%
}%
\begin{pgfscope}%
\pgfsys@transformshift{0.534489in}{0.645084in}%
\pgfsys@useobject{currentmarker}{}%
\end{pgfscope}%
\end{pgfscope}%
\begin{pgfscope}%
\pgfsetbuttcap%
\pgfsetroundjoin%
\definecolor{currentfill}{rgb}{0.000000,0.000000,0.000000}%
\pgfsetfillcolor{currentfill}%
\pgfsetlinewidth{0.602250pt}%
\definecolor{currentstroke}{rgb}{0.000000,0.000000,0.000000}%
\pgfsetstrokecolor{currentstroke}%
\pgfsetdash{}{0pt}%
\pgfsys@defobject{currentmarker}{\pgfqpoint{-0.027778in}{0.000000in}}{\pgfqpoint{-0.000000in}{0.000000in}}{%
\pgfpathmoveto{\pgfqpoint{-0.000000in}{0.000000in}}%
\pgfpathlineto{\pgfqpoint{-0.027778in}{0.000000in}}%
\pgfusepath{stroke,fill}%
}%
\begin{pgfscope}%
\pgfsys@transformshift{0.534489in}{0.721296in}%
\pgfsys@useobject{currentmarker}{}%
\end{pgfscope}%
\end{pgfscope}%
\begin{pgfscope}%
\pgfsetbuttcap%
\pgfsetroundjoin%
\definecolor{currentfill}{rgb}{0.000000,0.000000,0.000000}%
\pgfsetfillcolor{currentfill}%
\pgfsetlinewidth{0.602250pt}%
\definecolor{currentstroke}{rgb}{0.000000,0.000000,0.000000}%
\pgfsetstrokecolor{currentstroke}%
\pgfsetdash{}{0pt}%
\pgfsys@defobject{currentmarker}{\pgfqpoint{-0.027778in}{0.000000in}}{\pgfqpoint{-0.000000in}{0.000000in}}{%
\pgfpathmoveto{\pgfqpoint{-0.000000in}{0.000000in}}%
\pgfpathlineto{\pgfqpoint{-0.027778in}{0.000000in}}%
\pgfusepath{stroke,fill}%
}%
\begin{pgfscope}%
\pgfsys@transformshift{0.534489in}{0.873721in}%
\pgfsys@useobject{currentmarker}{}%
\end{pgfscope}%
\end{pgfscope}%
\begin{pgfscope}%
\pgfsetbuttcap%
\pgfsetroundjoin%
\definecolor{currentfill}{rgb}{0.000000,0.000000,0.000000}%
\pgfsetfillcolor{currentfill}%
\pgfsetlinewidth{0.602250pt}%
\definecolor{currentstroke}{rgb}{0.000000,0.000000,0.000000}%
\pgfsetstrokecolor{currentstroke}%
\pgfsetdash{}{0pt}%
\pgfsys@defobject{currentmarker}{\pgfqpoint{-0.027778in}{0.000000in}}{\pgfqpoint{-0.000000in}{0.000000in}}{%
\pgfpathmoveto{\pgfqpoint{-0.000000in}{0.000000in}}%
\pgfpathlineto{\pgfqpoint{-0.027778in}{0.000000in}}%
\pgfusepath{stroke,fill}%
}%
\begin{pgfscope}%
\pgfsys@transformshift{0.534489in}{0.949933in}%
\pgfsys@useobject{currentmarker}{}%
\end{pgfscope}%
\end{pgfscope}%
\begin{pgfscope}%
\pgfsetbuttcap%
\pgfsetroundjoin%
\definecolor{currentfill}{rgb}{0.000000,0.000000,0.000000}%
\pgfsetfillcolor{currentfill}%
\pgfsetlinewidth{0.602250pt}%
\definecolor{currentstroke}{rgb}{0.000000,0.000000,0.000000}%
\pgfsetstrokecolor{currentstroke}%
\pgfsetdash{}{0pt}%
\pgfsys@defobject{currentmarker}{\pgfqpoint{-0.027778in}{0.000000in}}{\pgfqpoint{-0.000000in}{0.000000in}}{%
\pgfpathmoveto{\pgfqpoint{-0.000000in}{0.000000in}}%
\pgfpathlineto{\pgfqpoint{-0.027778in}{0.000000in}}%
\pgfusepath{stroke,fill}%
}%
\begin{pgfscope}%
\pgfsys@transformshift{0.534489in}{1.026145in}%
\pgfsys@useobject{currentmarker}{}%
\end{pgfscope}%
\end{pgfscope}%
\begin{pgfscope}%
\pgfsetbuttcap%
\pgfsetroundjoin%
\definecolor{currentfill}{rgb}{0.000000,0.000000,0.000000}%
\pgfsetfillcolor{currentfill}%
\pgfsetlinewidth{0.602250pt}%
\definecolor{currentstroke}{rgb}{0.000000,0.000000,0.000000}%
\pgfsetstrokecolor{currentstroke}%
\pgfsetdash{}{0pt}%
\pgfsys@defobject{currentmarker}{\pgfqpoint{-0.027778in}{0.000000in}}{\pgfqpoint{-0.000000in}{0.000000in}}{%
\pgfpathmoveto{\pgfqpoint{-0.000000in}{0.000000in}}%
\pgfpathlineto{\pgfqpoint{-0.027778in}{0.000000in}}%
\pgfusepath{stroke,fill}%
}%
\begin{pgfscope}%
\pgfsys@transformshift{0.534489in}{1.102357in}%
\pgfsys@useobject{currentmarker}{}%
\end{pgfscope}%
\end{pgfscope}%
\begin{pgfscope}%
\pgfsetbuttcap%
\pgfsetroundjoin%
\definecolor{currentfill}{rgb}{0.000000,0.000000,0.000000}%
\pgfsetfillcolor{currentfill}%
\pgfsetlinewidth{0.602250pt}%
\definecolor{currentstroke}{rgb}{0.000000,0.000000,0.000000}%
\pgfsetstrokecolor{currentstroke}%
\pgfsetdash{}{0pt}%
\pgfsys@defobject{currentmarker}{\pgfqpoint{-0.027778in}{0.000000in}}{\pgfqpoint{-0.000000in}{0.000000in}}{%
\pgfpathmoveto{\pgfqpoint{-0.000000in}{0.000000in}}%
\pgfpathlineto{\pgfqpoint{-0.027778in}{0.000000in}}%
\pgfusepath{stroke,fill}%
}%
\begin{pgfscope}%
\pgfsys@transformshift{0.534489in}{1.254782in}%
\pgfsys@useobject{currentmarker}{}%
\end{pgfscope}%
\end{pgfscope}%
\begin{pgfscope}%
\pgfsetbuttcap%
\pgfsetroundjoin%
\definecolor{currentfill}{rgb}{0.000000,0.000000,0.000000}%
\pgfsetfillcolor{currentfill}%
\pgfsetlinewidth{0.602250pt}%
\definecolor{currentstroke}{rgb}{0.000000,0.000000,0.000000}%
\pgfsetstrokecolor{currentstroke}%
\pgfsetdash{}{0pt}%
\pgfsys@defobject{currentmarker}{\pgfqpoint{-0.027778in}{0.000000in}}{\pgfqpoint{-0.000000in}{0.000000in}}{%
\pgfpathmoveto{\pgfqpoint{-0.000000in}{0.000000in}}%
\pgfpathlineto{\pgfqpoint{-0.027778in}{0.000000in}}%
\pgfusepath{stroke,fill}%
}%
\begin{pgfscope}%
\pgfsys@transformshift{0.534489in}{1.330994in}%
\pgfsys@useobject{currentmarker}{}%
\end{pgfscope}%
\end{pgfscope}%
\begin{pgfscope}%
\pgfsetbuttcap%
\pgfsetroundjoin%
\definecolor{currentfill}{rgb}{0.000000,0.000000,0.000000}%
\pgfsetfillcolor{currentfill}%
\pgfsetlinewidth{0.602250pt}%
\definecolor{currentstroke}{rgb}{0.000000,0.000000,0.000000}%
\pgfsetstrokecolor{currentstroke}%
\pgfsetdash{}{0pt}%
\pgfsys@defobject{currentmarker}{\pgfqpoint{-0.027778in}{0.000000in}}{\pgfqpoint{-0.000000in}{0.000000in}}{%
\pgfpathmoveto{\pgfqpoint{-0.000000in}{0.000000in}}%
\pgfpathlineto{\pgfqpoint{-0.027778in}{0.000000in}}%
\pgfusepath{stroke,fill}%
}%
\begin{pgfscope}%
\pgfsys@transformshift{0.534489in}{1.407206in}%
\pgfsys@useobject{currentmarker}{}%
\end{pgfscope}%
\end{pgfscope}%
\begin{pgfscope}%
\pgfsetbuttcap%
\pgfsetroundjoin%
\definecolor{currentfill}{rgb}{0.000000,0.000000,0.000000}%
\pgfsetfillcolor{currentfill}%
\pgfsetlinewidth{0.602250pt}%
\definecolor{currentstroke}{rgb}{0.000000,0.000000,0.000000}%
\pgfsetstrokecolor{currentstroke}%
\pgfsetdash{}{0pt}%
\pgfsys@defobject{currentmarker}{\pgfqpoint{-0.027778in}{0.000000in}}{\pgfqpoint{-0.000000in}{0.000000in}}{%
\pgfpathmoveto{\pgfqpoint{-0.000000in}{0.000000in}}%
\pgfpathlineto{\pgfqpoint{-0.027778in}{0.000000in}}%
\pgfusepath{stroke,fill}%
}%
\begin{pgfscope}%
\pgfsys@transformshift{0.534489in}{1.483419in}%
\pgfsys@useobject{currentmarker}{}%
\end{pgfscope}%
\end{pgfscope}%
\begin{pgfscope}%
\pgfsetbuttcap%
\pgfsetroundjoin%
\definecolor{currentfill}{rgb}{0.000000,0.000000,0.000000}%
\pgfsetfillcolor{currentfill}%
\pgfsetlinewidth{0.602250pt}%
\definecolor{currentstroke}{rgb}{0.000000,0.000000,0.000000}%
\pgfsetstrokecolor{currentstroke}%
\pgfsetdash{}{0pt}%
\pgfsys@defobject{currentmarker}{\pgfqpoint{-0.027778in}{0.000000in}}{\pgfqpoint{-0.000000in}{0.000000in}}{%
\pgfpathmoveto{\pgfqpoint{-0.000000in}{0.000000in}}%
\pgfpathlineto{\pgfqpoint{-0.027778in}{0.000000in}}%
\pgfusepath{stroke,fill}%
}%
\begin{pgfscope}%
\pgfsys@transformshift{0.534489in}{1.635843in}%
\pgfsys@useobject{currentmarker}{}%
\end{pgfscope}%
\end{pgfscope}%
\begin{pgfscope}%
\pgfsetbuttcap%
\pgfsetroundjoin%
\definecolor{currentfill}{rgb}{0.000000,0.000000,0.000000}%
\pgfsetfillcolor{currentfill}%
\pgfsetlinewidth{0.602250pt}%
\definecolor{currentstroke}{rgb}{0.000000,0.000000,0.000000}%
\pgfsetstrokecolor{currentstroke}%
\pgfsetdash{}{0pt}%
\pgfsys@defobject{currentmarker}{\pgfqpoint{-0.027778in}{0.000000in}}{\pgfqpoint{-0.000000in}{0.000000in}}{%
\pgfpathmoveto{\pgfqpoint{-0.000000in}{0.000000in}}%
\pgfpathlineto{\pgfqpoint{-0.027778in}{0.000000in}}%
\pgfusepath{stroke,fill}%
}%
\begin{pgfscope}%
\pgfsys@transformshift{0.534489in}{1.712055in}%
\pgfsys@useobject{currentmarker}{}%
\end{pgfscope}%
\end{pgfscope}%
\begin{pgfscope}%
\pgfsetbuttcap%
\pgfsetroundjoin%
\definecolor{currentfill}{rgb}{0.000000,0.000000,0.000000}%
\pgfsetfillcolor{currentfill}%
\pgfsetlinewidth{0.602250pt}%
\definecolor{currentstroke}{rgb}{0.000000,0.000000,0.000000}%
\pgfsetstrokecolor{currentstroke}%
\pgfsetdash{}{0pt}%
\pgfsys@defobject{currentmarker}{\pgfqpoint{-0.027778in}{0.000000in}}{\pgfqpoint{-0.000000in}{0.000000in}}{%
\pgfpathmoveto{\pgfqpoint{-0.000000in}{0.000000in}}%
\pgfpathlineto{\pgfqpoint{-0.027778in}{0.000000in}}%
\pgfusepath{stroke,fill}%
}%
\begin{pgfscope}%
\pgfsys@transformshift{0.534489in}{1.788267in}%
\pgfsys@useobject{currentmarker}{}%
\end{pgfscope}%
\end{pgfscope}%
\begin{pgfscope}%
\pgfsetbuttcap%
\pgfsetroundjoin%
\definecolor{currentfill}{rgb}{0.000000,0.000000,0.000000}%
\pgfsetfillcolor{currentfill}%
\pgfsetlinewidth{0.602250pt}%
\definecolor{currentstroke}{rgb}{0.000000,0.000000,0.000000}%
\pgfsetstrokecolor{currentstroke}%
\pgfsetdash{}{0pt}%
\pgfsys@defobject{currentmarker}{\pgfqpoint{-0.027778in}{0.000000in}}{\pgfqpoint{-0.000000in}{0.000000in}}{%
\pgfpathmoveto{\pgfqpoint{-0.000000in}{0.000000in}}%
\pgfpathlineto{\pgfqpoint{-0.027778in}{0.000000in}}%
\pgfusepath{stroke,fill}%
}%
\begin{pgfscope}%
\pgfsys@transformshift{0.534489in}{1.864480in}%
\pgfsys@useobject{currentmarker}{}%
\end{pgfscope}%
\end{pgfscope}%
\begin{pgfscope}%
\definecolor{textcolor}{rgb}{0.000000,0.000000,0.000000}%
\pgfsetstrokecolor{textcolor}%
\pgfsetfillcolor{textcolor}%
\pgftext[x=0.177211in,y=1.178570in,,bottom,rotate=90.000000]{\color{textcolor}\rmfamily\fontsize{9.000000}{10.800000}\selectfont \(\displaystyle \Phi_{\bm{k}}(t)/\Phi_{\bm{k}}^{\mathrm{ext}}(t)\)}%
\end{pgfscope}%
\begin{pgfscope}%
\pgfpathrectangle{\pgfqpoint{0.534489in}{0.416447in}}{\pgfqpoint{2.717944in}{1.524244in}}%
\pgfusepath{clip}%
\pgfsetrectcap%
\pgfsetroundjoin%
\pgfsetlinewidth{0.803000pt}%
\definecolor{currentstroke}{rgb}{0.690196,0.690196,0.690196}%
\pgfsetstrokecolor{currentstroke}%
\pgfsetdash{}{0pt}%
\pgfpathmoveto{\pgfqpoint{1.893461in}{0.416447in}}%
\pgfpathlineto{\pgfqpoint{1.893461in}{1.940692in}}%
\pgfusepath{stroke}%
\end{pgfscope}%
\begin{pgfscope}%
\pgfpathrectangle{\pgfqpoint{0.534489in}{0.416447in}}{\pgfqpoint{2.717944in}{1.524244in}}%
\pgfusepath{clip}%
\pgfsetbuttcap%
\pgfsetroundjoin%
\pgfsetlinewidth{0.803000pt}%
\definecolor{currentstroke}{rgb}{0.690196,0.690196,0.690196}%
\pgfsetstrokecolor{currentstroke}%
\pgfsetdash{{2.960000pt}{1.280000pt}}{0.000000pt}%
\pgfpathmoveto{\pgfqpoint{0.534489in}{1.178570in}}%
\pgfpathlineto{\pgfqpoint{3.252433in}{1.178570in}}%
\pgfusepath{stroke}%
\end{pgfscope}%
\begin{pgfscope}%
\pgfpathrectangle{\pgfqpoint{0.534489in}{0.416447in}}{\pgfqpoint{2.717944in}{1.524244in}}%
\pgfusepath{clip}%
\pgfsetrectcap%
\pgfsetroundjoin%
\pgfsetlinewidth{1.505625pt}%
\definecolor{currentstroke}{rgb}{0.121569,0.466667,0.705882}%
\pgfsetstrokecolor{currentstroke}%
\pgfsetdash{}{0pt}%
\pgfpathmoveto{\pgfqpoint{0.524489in}{1.229131in}}%
\pgfpathlineto{\pgfqpoint{0.732262in}{1.235280in}}%
\pgfpathlineto{\pgfqpoint{0.900264in}{1.242366in}}%
\pgfpathlineto{\pgfqpoint{1.031517in}{1.249971in}}%
\pgfpathlineto{\pgfqpoint{1.141769in}{1.258516in}}%
\pgfpathlineto{\pgfqpoint{1.231020in}{1.267592in}}%
\pgfpathlineto{\pgfqpoint{1.304522in}{1.277183in}}%
\pgfpathlineto{\pgfqpoint{1.367523in}{1.287579in}}%
\pgfpathlineto{\pgfqpoint{1.420024in}{1.298386in}}%
\pgfpathlineto{\pgfqpoint{1.467274in}{1.310415in}}%
\pgfpathlineto{\pgfqpoint{1.504025in}{1.321806in}}%
\pgfpathlineto{\pgfqpoint{1.540776in}{1.335588in}}%
\pgfpathlineto{\pgfqpoint{1.572276in}{1.349926in}}%
\pgfpathlineto{\pgfqpoint{1.598527in}{1.364225in}}%
\pgfpathlineto{\pgfqpoint{1.624777in}{1.381328in}}%
\pgfpathlineto{\pgfqpoint{1.645778in}{1.397629in}}%
\pgfpathlineto{\pgfqpoint{1.666778in}{1.416958in}}%
\pgfpathlineto{\pgfqpoint{1.682528in}{1.433986in}}%
\pgfpathlineto{\pgfqpoint{1.698278in}{1.453768in}}%
\pgfpathlineto{\pgfqpoint{1.714029in}{1.477027in}}%
\pgfpathlineto{\pgfqpoint{1.729779in}{1.504769in}}%
\pgfpathlineto{\pgfqpoint{1.745529in}{1.538424in}}%
\pgfpathlineto{\pgfqpoint{1.756029in}{1.565151in}}%
\pgfpathlineto{\pgfqpoint{1.766530in}{1.596302in}}%
\pgfpathlineto{\pgfqpoint{1.777030in}{1.633076in}}%
\pgfpathlineto{\pgfqpoint{1.787530in}{1.677144in}}%
\pgfpathlineto{\pgfqpoint{1.798030in}{1.730913in}}%
\pgfpathlineto{\pgfqpoint{1.808530in}{1.797983in}}%
\pgfpathlineto{\pgfqpoint{1.819031in}{1.883982in}}%
\pgfpathlineto{\pgfqpoint{1.825470in}{1.950692in}}%
\pgfpathlineto{\pgfqpoint{1.825470in}{1.950692in}}%
\pgfusepath{stroke}%
\end{pgfscope}%
\begin{pgfscope}%
\pgfpathrectangle{\pgfqpoint{0.534489in}{0.416447in}}{\pgfqpoint{2.717944in}{1.524244in}}%
\pgfusepath{clip}%
\pgfsetrectcap%
\pgfsetroundjoin%
\pgfsetlinewidth{1.505625pt}%
\definecolor{currentstroke}{rgb}{0.121569,0.466667,0.705882}%
\pgfsetstrokecolor{currentstroke}%
\pgfsetdash{}{0pt}%
\pgfpathmoveto{\pgfqpoint{1.959824in}{0.406447in}}%
\pgfpathlineto{\pgfqpoint{1.968031in}{0.493523in}}%
\pgfpathlineto{\pgfqpoint{1.978563in}{0.579442in}}%
\pgfpathlineto{\pgfqpoint{1.989094in}{0.646435in}}%
\pgfpathlineto{\pgfqpoint{1.999626in}{0.700133in}}%
\pgfpathlineto{\pgfqpoint{2.010157in}{0.744135in}}%
\pgfpathlineto{\pgfqpoint{2.020688in}{0.780850in}}%
\pgfpathlineto{\pgfqpoint{2.036485in}{0.825780in}}%
\pgfpathlineto{\pgfqpoint{2.052282in}{0.861767in}}%
\pgfpathlineto{\pgfqpoint{2.068079in}{0.891238in}}%
\pgfpathlineto{\pgfqpoint{2.083876in}{0.915815in}}%
\pgfpathlineto{\pgfqpoint{2.099674in}{0.936623in}}%
\pgfpathlineto{\pgfqpoint{2.115471in}{0.954466in}}%
\pgfpathlineto{\pgfqpoint{2.136533in}{0.974644in}}%
\pgfpathlineto{\pgfqpoint{2.157596in}{0.991599in}}%
\pgfpathlineto{\pgfqpoint{2.178659in}{1.006045in}}%
\pgfpathlineto{\pgfqpoint{2.204987in}{1.021349in}}%
\pgfpathlineto{\pgfqpoint{2.231316in}{1.034262in}}%
\pgfpathlineto{\pgfqpoint{2.262910in}{1.047321in}}%
\pgfpathlineto{\pgfqpoint{2.299770in}{1.059981in}}%
\pgfpathlineto{\pgfqpoint{2.341895in}{1.071890in}}%
\pgfpathlineto{\pgfqpoint{2.389286in}{1.082858in}}%
\pgfpathlineto{\pgfqpoint{2.441943in}{1.092808in}}%
\pgfpathlineto{\pgfqpoint{2.505131in}{1.102471in}}%
\pgfpathlineto{\pgfqpoint{2.578851in}{1.111474in}}%
\pgfpathlineto{\pgfqpoint{2.668368in}{1.120080in}}%
\pgfpathlineto{\pgfqpoint{2.773681in}{1.127937in}}%
\pgfpathlineto{\pgfqpoint{2.905323in}{1.135423in}}%
\pgfpathlineto{\pgfqpoint{3.068560in}{1.142334in}}%
\pgfpathlineto{\pgfqpoint{3.262433in}{1.148352in}}%
\pgfpathlineto{\pgfqpoint{3.262433in}{1.148352in}}%
\pgfusepath{stroke}%
\end{pgfscope}%
\begin{pgfscope}%
\pgfpathrectangle{\pgfqpoint{0.534489in}{0.416447in}}{\pgfqpoint{2.717944in}{1.524244in}}%
\pgfusepath{clip}%
\pgfsetrectcap%
\pgfsetroundjoin%
\pgfsetlinewidth{1.505625pt}%
\definecolor{currentstroke}{rgb}{1.000000,0.498039,0.054902}%
\pgfsetstrokecolor{currentstroke}%
\pgfsetdash{}{0pt}%
\pgfpathmoveto{\pgfqpoint{0.524489in}{1.178568in}}%
\pgfpathlineto{\pgfqpoint{1.068267in}{1.177471in}}%
\pgfpathlineto{\pgfqpoint{1.210020in}{1.175084in}}%
\pgfpathlineto{\pgfqpoint{1.304522in}{1.171470in}}%
\pgfpathlineto{\pgfqpoint{1.378023in}{1.166474in}}%
\pgfpathlineto{\pgfqpoint{1.435774in}{1.160312in}}%
\pgfpathlineto{\pgfqpoint{1.483025in}{1.153018in}}%
\pgfpathlineto{\pgfqpoint{1.519775in}{1.145324in}}%
\pgfpathlineto{\pgfqpoint{1.551276in}{1.136790in}}%
\pgfpathlineto{\pgfqpoint{1.582776in}{1.125841in}}%
\pgfpathlineto{\pgfqpoint{1.609027in}{1.114252in}}%
\pgfpathlineto{\pgfqpoint{1.630027in}{1.102861in}}%
\pgfpathlineto{\pgfqpoint{1.651028in}{1.089033in}}%
\pgfpathlineto{\pgfqpoint{1.672028in}{1.072054in}}%
\pgfpathlineto{\pgfqpoint{1.687778in}{1.056657in}}%
\pgfpathlineto{\pgfqpoint{1.703529in}{1.038334in}}%
\pgfpathlineto{\pgfqpoint{1.719279in}{1.016280in}}%
\pgfpathlineto{\pgfqpoint{1.735029in}{0.989369in}}%
\pgfpathlineto{\pgfqpoint{1.745529in}{0.967956in}}%
\pgfpathlineto{\pgfqpoint{1.756029in}{0.943018in}}%
\pgfpathlineto{\pgfqpoint{1.766530in}{0.913675in}}%
\pgfpathlineto{\pgfqpoint{1.777030in}{0.878730in}}%
\pgfpathlineto{\pgfqpoint{1.787530in}{0.836510in}}%
\pgfpathlineto{\pgfqpoint{1.798030in}{0.784609in}}%
\pgfpathlineto{\pgfqpoint{1.808530in}{0.719428in}}%
\pgfpathlineto{\pgfqpoint{1.819031in}{0.635337in}}%
\pgfpathlineto{\pgfqpoint{1.829531in}{0.523011in}}%
\pgfpathlineto{\pgfqpoint{1.837541in}{0.406447in}}%
\pgfpathlineto{\pgfqpoint{1.837541in}{0.406447in}}%
\pgfusepath{stroke}%
\end{pgfscope}%
\begin{pgfscope}%
\pgfpathrectangle{\pgfqpoint{0.534489in}{0.416447in}}{\pgfqpoint{2.717944in}{1.524244in}}%
\pgfusepath{clip}%
\pgfsetrectcap%
\pgfsetroundjoin%
\pgfsetlinewidth{1.505625pt}%
\definecolor{currentstroke}{rgb}{1.000000,0.498039,0.054902}%
\pgfsetstrokecolor{currentstroke}%
\pgfsetdash{}{0pt}%
\pgfpathmoveto{\pgfqpoint{1.981726in}{1.950692in}}%
\pgfpathlineto{\pgfqpoint{1.989094in}{1.906563in}}%
\pgfpathlineto{\pgfqpoint{1.999626in}{1.855120in}}%
\pgfpathlineto{\pgfqpoint{2.010157in}{1.813392in}}%
\pgfpathlineto{\pgfqpoint{2.020688in}{1.778971in}}%
\pgfpathlineto{\pgfqpoint{2.031220in}{1.750185in}}%
\pgfpathlineto{\pgfqpoint{2.047017in}{1.715046in}}%
\pgfpathlineto{\pgfqpoint{2.062814in}{1.687164in}}%
\pgfpathlineto{\pgfqpoint{2.078611in}{1.664692in}}%
\pgfpathlineto{\pgfqpoint{2.094408in}{1.646365in}}%
\pgfpathlineto{\pgfqpoint{2.110205in}{1.631285in}}%
\pgfpathlineto{\pgfqpoint{2.126002in}{1.618799in}}%
\pgfpathlineto{\pgfqpoint{2.147065in}{1.605361in}}%
\pgfpathlineto{\pgfqpoint{2.168127in}{1.594832in}}%
\pgfpathlineto{\pgfqpoint{2.189190in}{1.586607in}}%
\pgfpathlineto{\pgfqpoint{2.215519in}{1.578895in}}%
\pgfpathlineto{\pgfqpoint{2.241847in}{1.573457in}}%
\pgfpathlineto{\pgfqpoint{2.273441in}{1.569297in}}%
\pgfpathlineto{\pgfqpoint{2.310301in}{1.567004in}}%
\pgfpathlineto{\pgfqpoint{2.352426in}{1.567001in}}%
\pgfpathlineto{\pgfqpoint{2.399818in}{1.569572in}}%
\pgfpathlineto{\pgfqpoint{2.452474in}{1.574895in}}%
\pgfpathlineto{\pgfqpoint{2.510397in}{1.583075in}}%
\pgfpathlineto{\pgfqpoint{2.578851in}{1.595180in}}%
\pgfpathlineto{\pgfqpoint{2.652570in}{1.610522in}}%
\pgfpathlineto{\pgfqpoint{2.736821in}{1.630343in}}%
\pgfpathlineto{\pgfqpoint{2.831604in}{1.654915in}}%
\pgfpathlineto{\pgfqpoint{2.942183in}{1.685979in}}%
\pgfpathlineto{\pgfqpoint{3.068560in}{1.723888in}}%
\pgfpathlineto{\pgfqpoint{3.215999in}{1.770527in}}%
\pgfpathlineto{\pgfqpoint{3.262433in}{1.785633in}}%
\pgfpathlineto{\pgfqpoint{3.262433in}{1.785633in}}%
\pgfusepath{stroke}%
\end{pgfscope}%
\begin{pgfscope}%
\pgfsetrectcap%
\pgfsetmiterjoin%
\pgfsetlinewidth{0.803000pt}%
\definecolor{currentstroke}{rgb}{0.000000,0.000000,0.000000}%
\pgfsetstrokecolor{currentstroke}%
\pgfsetdash{}{0pt}%
\pgfpathmoveto{\pgfqpoint{0.534489in}{0.416447in}}%
\pgfpathlineto{\pgfqpoint{0.534489in}{1.940692in}}%
\pgfusepath{stroke}%
\end{pgfscope}%
\begin{pgfscope}%
\pgfsetrectcap%
\pgfsetmiterjoin%
\pgfsetlinewidth{0.803000pt}%
\definecolor{currentstroke}{rgb}{0.000000,0.000000,0.000000}%
\pgfsetstrokecolor{currentstroke}%
\pgfsetdash{}{0pt}%
\pgfpathmoveto{\pgfqpoint{3.252433in}{0.416447in}}%
\pgfpathlineto{\pgfqpoint{3.252433in}{1.940692in}}%
\pgfusepath{stroke}%
\end{pgfscope}%
\begin{pgfscope}%
\pgfsetrectcap%
\pgfsetmiterjoin%
\pgfsetlinewidth{0.803000pt}%
\definecolor{currentstroke}{rgb}{0.000000,0.000000,0.000000}%
\pgfsetstrokecolor{currentstroke}%
\pgfsetdash{}{0pt}%
\pgfpathmoveto{\pgfqpoint{0.534489in}{0.416447in}}%
\pgfpathlineto{\pgfqpoint{3.252433in}{0.416447in}}%
\pgfusepath{stroke}%
\end{pgfscope}%
\begin{pgfscope}%
\pgfsetrectcap%
\pgfsetmiterjoin%
\pgfsetlinewidth{0.803000pt}%
\definecolor{currentstroke}{rgb}{0.000000,0.000000,0.000000}%
\pgfsetstrokecolor{currentstroke}%
\pgfsetdash{}{0pt}%
\pgfpathmoveto{\pgfqpoint{0.534489in}{1.940692in}}%
\pgfpathlineto{\pgfqpoint{3.252433in}{1.940692in}}%
\pgfusepath{stroke}%
\end{pgfscope}%
\begin{pgfscope}%
\pgfpathrectangle{\pgfqpoint{0.534489in}{0.416447in}}{\pgfqpoint{2.717944in}{1.524244in}}%
\pgfusepath{clip}%
\pgfsetrectcap%
\pgfsetroundjoin%
\pgfsetlinewidth{1.505625pt}%
\definecolor{currentstroke}{rgb}{0.000000,0.000000,0.000000}%
\pgfsetstrokecolor{currentstroke}%
\pgfsetdash{}{0pt}%
\pgfpathmoveto{\pgfqpoint{0.532758in}{1.229333in}}%
\pgfpathlineto{\pgfqpoint{0.748012in}{1.235807in}}%
\pgfpathlineto{\pgfqpoint{0.921265in}{1.243155in}}%
\pgfpathlineto{\pgfqpoint{1.073518in}{1.251796in}}%
\pgfpathlineto{\pgfqpoint{1.210020in}{1.261752in}}%
\pgfpathlineto{\pgfqpoint{1.336022in}{1.273147in}}%
\pgfpathlineto{\pgfqpoint{1.456774in}{1.286318in}}%
\pgfpathlineto{\pgfqpoint{1.572276in}{1.301162in}}%
\pgfpathlineto{\pgfqpoint{1.682528in}{1.317498in}}%
\pgfpathlineto{\pgfqpoint{1.792780in}{1.336016in}}%
\pgfpathlineto{\pgfqpoint{1.861031in}{1.348584in}}%
\pgfpathlineto{\pgfqpoint{2.004891in}{1.377818in}}%
\pgfpathlineto{\pgfqpoint{2.120736in}{1.403998in}}%
\pgfpathlineto{\pgfqpoint{2.236581in}{1.432423in}}%
\pgfpathlineto{\pgfqpoint{2.357692in}{1.464391in}}%
\pgfpathlineto{\pgfqpoint{2.484069in}{1.500003in}}%
\pgfpathlineto{\pgfqpoint{2.620976in}{1.540911in}}%
\pgfpathlineto{\pgfqpoint{2.768416in}{1.587317in}}%
\pgfpathlineto{\pgfqpoint{2.931652in}{1.641045in}}%
\pgfpathlineto{\pgfqpoint{3.121216in}{1.705879in}}%
\pgfpathlineto{\pgfqpoint{3.252859in}{1.752030in}}%
\pgfpathlineto{\pgfqpoint{3.252859in}{1.752030in}}%
\pgfusepath{stroke}%
\end{pgfscope}%
\begin{pgfscope}%
\pgfsetbuttcap%
\pgfsetmiterjoin%
\definecolor{currentfill}{rgb}{1.000000,1.000000,1.000000}%
\pgfsetfillcolor{currentfill}%
\pgfsetlinewidth{1.003750pt}%
\definecolor{currentstroke}{rgb}{0.800000,0.800000,0.800000}%
\pgfsetstrokecolor{currentstroke}%
\pgfsetdash{}{0pt}%
\pgfpathmoveto{\pgfqpoint{0.612266in}{0.472003in}}%
\pgfpathlineto{\pgfqpoint{1.348044in}{0.472003in}}%
\pgfpathquadraticcurveto{\pgfqpoint{1.370266in}{0.472003in}}{\pgfqpoint{1.370266in}{0.494225in}}%
\pgfpathlineto{\pgfqpoint{1.370266in}{1.064571in}}%
\pgfpathquadraticcurveto{\pgfqpoint{1.370266in}{1.086793in}}{\pgfqpoint{1.348044in}{1.086793in}}%
\pgfpathlineto{\pgfqpoint{0.612266in}{1.086793in}}%
\pgfpathquadraticcurveto{\pgfqpoint{0.590044in}{1.086793in}}{\pgfqpoint{0.590044in}{1.064571in}}%
\pgfpathlineto{\pgfqpoint{0.590044in}{0.494225in}}%
\pgfpathquadraticcurveto{\pgfqpoint{0.590044in}{0.472003in}}{\pgfqpoint{0.612266in}{0.472003in}}%
\pgfpathlineto{\pgfqpoint{0.612266in}{0.472003in}}%
\pgfpathclose%
\pgfusepath{stroke,fill}%
\end{pgfscope}%
\begin{pgfscope}%
\pgfsetrectcap%
\pgfsetroundjoin%
\pgfsetlinewidth{1.505625pt}%
\definecolor{currentstroke}{rgb}{0.000000,0.000000,0.000000}%
\pgfsetstrokecolor{currentstroke}%
\pgfsetdash{}{0pt}%
\pgfpathmoveto{\pgfqpoint{0.634489in}{0.987098in}}%
\pgfpathlineto{\pgfqpoint{0.745600in}{0.987098in}}%
\pgfpathlineto{\pgfqpoint{0.856711in}{0.987098in}}%
\pgfusepath{stroke}%
\end{pgfscope}%
\begin{pgfscope}%
\definecolor{textcolor}{rgb}{0.000000,0.000000,0.000000}%
\pgfsetstrokecolor{textcolor}%
\pgfsetfillcolor{textcolor}%
\pgftext[x=0.945600in,y=0.948209in,left,base]{\color{textcolor}\rmfamily\fontsize{8.000000}{9.600000}\selectfont \(\displaystyle \Phi_{\bm{k}}^{\mathrm{tot}}(t)\)}%
\end{pgfscope}%
\begin{pgfscope}%
\pgfsetrectcap%
\pgfsetroundjoin%
\pgfsetlinewidth{1.505625pt}%
\definecolor{currentstroke}{rgb}{0.121569,0.466667,0.705882}%
\pgfsetstrokecolor{currentstroke}%
\pgfsetdash{}{0pt}%
\pgfpathmoveto{\pgfqpoint{0.634489in}{0.790488in}}%
\pgfpathlineto{\pgfqpoint{0.745600in}{0.790488in}}%
\pgfpathlineto{\pgfqpoint{0.856711in}{0.790488in}}%
\pgfusepath{stroke}%
\end{pgfscope}%
\begin{pgfscope}%
\definecolor{textcolor}{rgb}{0.000000,0.000000,0.000000}%
\pgfsetstrokecolor{textcolor}%
\pgfsetfillcolor{textcolor}%
\pgftext[x=0.945600in,y=0.751599in,left,base]{\color{textcolor}\rmfamily\fontsize{8.000000}{9.600000}\selectfont \(\displaystyle \Phi_{\bm{k}}^{\mathrm{wake}}(t)\)}%
\end{pgfscope}%
\begin{pgfscope}%
\pgfsetrectcap%
\pgfsetroundjoin%
\pgfsetlinewidth{1.505625pt}%
\definecolor{currentstroke}{rgb}{1.000000,0.498039,0.054902}%
\pgfsetstrokecolor{currentstroke}%
\pgfsetdash{}{0pt}%
\pgfpathmoveto{\pgfqpoint{0.634489in}{0.593627in}}%
\pgfpathlineto{\pgfqpoint{0.745600in}{0.593627in}}%
\pgfpathlineto{\pgfqpoint{0.856711in}{0.593627in}}%
\pgfusepath{stroke}%
\end{pgfscope}%
\begin{pgfscope}%
\definecolor{textcolor}{rgb}{0.000000,0.000000,0.000000}%
\pgfsetstrokecolor{textcolor}%
\pgfsetfillcolor{textcolor}%
\pgftext[x=0.945600in,y=0.554739in,left,base]{\color{textcolor}\rmfamily\fontsize{8.000000}{9.600000}\selectfont \(\displaystyle \Phi_{\bm{k}}^{\mathrm{mode}}(t)\)}%
\end{pgfscope}%
\end{pgfpicture}%
\makeatother%
\endgroup%

%% file: paper.bbl
\begin{thebibliography}{}
\makeatletter
\relax
\def\mn@urlcharsother{\let\do\@makeother \do\$\do\&\do\#\do\^\do\_\do\%\do\~}
\def\mn@doi{\begingroup\mn@urlcharsother \@ifnextchar [ {\mn@doi@}
  {\mn@doi@[]}}
\def\mn@doi@[#1]#2{\def\@tempa{#1}\ifx\@tempa\@empty \href
  {http://dx.doi.org/#2} {doi:#2}\else \href {http://dx.doi.org/#2} {#1}\fi
  \endgroup}
\def\mn@eprint#1#2{\mn@eprint@#1:#2::\@nil}
\def\mn@eprint@arXiv#1{\href {http://arxiv.org/abs/#1} {{\tt arXiv:#1}}}
\def\mn@eprint@dblp#1{\href {http://dblp.uni-trier.de/rec/bibtex/#1.xml}
  {dblp:#1}}
\def\mn@eprint@#1:#2:#3:#4\@nil{\def\@tempa {#1}\def\@tempb {#2}\def\@tempc
  {#3}\ifx \@tempc \@empty \let \@tempc \@tempb \let \@tempb \@tempa \fi \ifx
  \@tempb \@empty \def\@tempb {arXiv}\fi \@ifundefined
  {mn@eprint@\@tempb}{\@tempb:\@tempc}{\expandafter \expandafter \csname
  mn@eprint@\@tempb\endcsname \expandafter{\@tempc}}}

\bibitem[\protect\citeauthoryear{B{\u a}lescu}{B{\u
  a}lescu}{1963}]{Balescu1963-ye}
B{\u a}lescu R.,  1963, Journal of mathematical physics, 4, 1009

\bibitem[\protect\citeauthoryear{Begelman, Blandford  \& Rees}{Begelman
  et~al.}{1980}]{begelman1980massive}
Begelman M.~C.,  Blandford R.~D.,   Rees M.~J.,  1980, Nature, 287, 307

\bibitem[\protect\citeauthoryear{Binney}{Binney}{2020}]{binney2020shearing}
Binney J.,  2020, Monthly Notices of the Royal Astronomical Society, 496, 767

\bibitem[\protect\citeauthoryear{Binney \& Lacey}{Binney \&
  Lacey}{1988}]{Binney1988-zy}
Binney J.,  Lacey C.,  1988, Monthly Notices of the Royal Astronomical Society,
  230, 597

\bibitem[\protect\citeauthoryear{Binney \& Tremaine}{Binney \&
  Tremaine}{2008}]{Binney2008-ou}
Binney J.,  Tremaine S.,  2008, Galactic Dynamics: Second Edition, by James
  Binney and Scott Tremaine. ISBN 978-0-691-13026-2 (HB). Published by
  Princeton University Press, Princeton, NJ USA, 2008.

\bibitem[\protect\citeauthoryear{Boylan-Kolchin, Ma  \&
  Quataert}{Boylan-Kolchin et~al.}{2008}]{boylan2008dynamical}
Boylan-Kolchin M.,  Ma C.-P.,   Quataert E.,  2008, Monthly Notices of the
  Royal Astronomical Society, 383, 93

\bibitem[\protect\citeauthoryear{Bălescu}{Bălescu}{1963}]{Balescu1963}
Bălescu R.,  1963, Statistical {{Mechanics}} of {{Charged Particles}}.
 Monographs in {{Statistical Physics}} and {{Thermodynamics}} Vol. 4,
  {Interscience Publishers}

\bibitem[\protect\citeauthoryear{Chandrasekhar}{Chandrasekhar}{1943}]{chandrasekhar1943dynamical}
Chandrasekhar S.,  1943, Astrophysical Journal, 97, 255

\bibitem[\protect\citeauthoryear{Chavanis}{Chavanis}{2012}]{chavanis2012kinetic}
Chavanis P.-H.,  2012, Physica A: Statistical Mechanics and its Applications,
  391, 3680

\bibitem[\protect\citeauthoryear{Chavanis}{Chavanis}{2023}]{chavanis2023secular}
Chavanis P.-H.,  2023, Universe, 9, 68

\bibitem[\protect\citeauthoryear{D’Onghia, Vogelsberger  \&
  Hernquist}{D’Onghia et~al.}{2013}]{d2013self}
D’Onghia E.,  Vogelsberger M.,   Hernquist L.,  2013, The Astrophysical
  Journal, 766, 34

\bibitem[\protect\citeauthoryear{Fouvry \& Bar-Or}{Fouvry \&
  Bar-Or}{2018}]{Fouvry2018-gi}
Fouvry J.-B.,  Bar-Or B.,  2018, Monthly Notices of the Royal Astronomical
  Society, 481, 4566

\bibitem[\protect\citeauthoryear{Fouvry, Pichon, Magorrian  \& Chavanis}{Fouvry
  et~al.}{2015}]{Fouvry2015-nk}
Fouvry J.~B.,  Pichon C.,  Magorrian J.,   Chavanis P.~H.,  2015, Astronomy \&
  Astrophysics, 584, A129

\bibitem[\protect\citeauthoryear{Fouvry, Bar-Or  \& Chavanis}{Fouvry
  et~al.}{2019}]{fouvry2019secular}
Fouvry J.-B.,  Bar-Or B.,   Chavanis P.-H.,  2019, Physical Review E, 99,
  032101

\bibitem[\protect\citeauthoryear{Hamilton}{Hamilton}{2021}]{Hamilton2021-qe}
Hamilton C.,  2021, Monthly Notices of the Royal Astronomical Society

\bibitem[\protect\citeauthoryear{Hatori}{Hatori}{1969}]{hatori1969nonstationary}
Hatori T.,  1969, The Physics of Fluids, 12, 1652

\bibitem[\protect\citeauthoryear{Heyvaerts}{Heyvaerts}{2010}]{heyvaerts2010balescu}
Heyvaerts J.,  2010, Monthly Notices of the Royal Astronomical Society, 407,
  355

\bibitem[\protect\citeauthoryear{Ichimaru}{Ichimaru}{1973}]{Ichimaru1973}
Ichimaru S.,  1973, Basic Principles of Plasma Physics: A Statistical Approach.
Frontiers in Physics, {W. A. Benjamin, Inc.}

\bibitem[\protect\citeauthoryear{Julian \& Toomre}{Julian \&
  Toomre}{1966}]{julian1966non}
Julian W.~H.,  Toomre A.,  1966, The Astrophysical Journal, 146, 810

\bibitem[\protect\citeauthoryear{Lancaster, Giovanetti, Mocz, Kahn, Lisanti  \&
  Spergel}{Lancaster et~al.}{2020}]{Lancaster2020-vt}
Lancaster L.,  Giovanetti C.,  Mocz P.,  Kahn Y.,  Lisanti M.,   Spergel D.~N.,
   2020, Journal of Cosmology and Astroparticle Physics, 2020, 001

\bibitem[\protect\citeauthoryear{Lee \& Shadwick}{Lee \&
  Shadwick}{2023}]{lee2023cauchy}
Lee F.~M.,  Shadwick B.,  2023, Physical Review E, 107, L063201

\bibitem[\protect\citeauthoryear{Lin \& Shu}{Lin \& Shu}{1964}]{lin1964spiral}
Lin C.,  Shu F.~H.,  1964, The Astrophysical Journal, 140, 646

\bibitem[\protect\citeauthoryear{Lynden-Bell \& Kalnajs}{Lynden-Bell \&
  Kalnajs}{1972}]{Lynden-Bell1972-ve}
Lynden-Bell D.,  Kalnajs A.~J.,  1972, Monthly Notices of the Royal
  Astronomical Society, 157, 1

\bibitem[\protect\citeauthoryear{Magorrian}{Magorrian}{2021}]{magorrian2021stellar}
Magorrian J.,  2021, Monthly Notices of the Royal Astronomical Society, 507,
  4840

\bibitem[\protect\citeauthoryear{Nelson \& Tremaine}{Nelson \&
  Tremaine}{1999}]{Nelson1999-in}
Nelson R.~W.,  Tremaine S.,  1999, Monthly Notices of the Royal Astronomical
  Society, 306, 1

\bibitem[\protect\citeauthoryear{Oberman}{Oberman}{1970}]{oberman1970advanced}
Oberman C.,  1970, Physics of Hot Plasmas: Scottish Universities’ Summer
  School 1968, pp 42--102

\bibitem[\protect\citeauthoryear{Palmer}{Palmer}{1994}]{palmer1994stability}
Palmer P.,  1994, Astrophysics and Space Science Library, 185

\bibitem[\protect\citeauthoryear{Pichon \& Aubert}{Pichon \&
  Aubert}{2006}]{Pichon2006-ak}
Pichon C.,  Aubert D.,  2006, Monthly Notices of the Royal Astronomical
  Society, 368, 1657

\bibitem[\protect\citeauthoryear{Rogister \& Oberman}{Rogister \&
  Oberman}{1968}]{Rogister1968-tb}
Rogister A.,  Oberman C.,  1968, Journal of Plasma Physics, 2, 33

\bibitem[\protect\citeauthoryear{Sellwood}{Sellwood}{2021}]{Sellwood2021-rt}
Sellwood J.~A.,  2021, Monthly Notices of the Royal Astronomical Society, 506,
  3018

\bibitem[\protect\citeauthoryear{Sellwood \& Carlberg}{Sellwood \&
  Carlberg}{2014}]{Sellwood2014-xo}
Sellwood J.~A.,  Carlberg R.~G.,  2014, The Astrophysical Journal, 785, 137

\bibitem[\protect\citeauthoryear{Thorne \& Blandford}{Thorne \&
  Blandford}{2017}]{thorne2017modern}
Thorne K.~S.,  Blandford R.~D.,  2017, Modern classical physics: optics,
  fluids, plasmas, elasticity, relativity, and statistical physics.
Princeton University Press

\bibitem[\protect\citeauthoryear{Tremaine \& Weinberg}{Tremaine \&
  Weinberg}{1984}]{Tremaine1984-wt}
Tremaine S.,  Weinberg M.~D.,  1984, Monthly Notices of the Royal Astronomical
  Society, 209, 729

\bibitem[\protect\citeauthoryear{Weinberg}{Weinberg}{1985}]{Weinberg1985-en}
Weinberg M.~D.,  1985, Monthly Notices of the Royal Astronomical Society, 213,
  451

\bibitem[\protect\citeauthoryear{Weinberg}{Weinberg}{1989}]{Weinberg1989-cj}
Weinberg M.~D.,  1989, Monthly Notices of the Royal Astronomical Society, 239,
  549

\bibitem[\protect\citeauthoryear{Weinberg}{Weinberg}{1993}]{weinberg1993nonlocal}
Weinberg M.~D.,  1993, The Astrophysical Journal, 410, 543

\bibitem[\protect\citeauthoryear{Weinberg}{Weinberg}{2001}]{Weinberg2001-ok}
Weinberg M.~D.,  2001, Monthly Notices of the Royal Astronomical Society, 328,
  311

\makeatother
\end{thebibliography}
